\documentstyle[preprint,tighten,aps,epsf]{revtex}
\def\psibar{\overline{\psi}}
\def\csw{c_{\rm sw}}
\def\bm{b_{\rm m}}
\def\kcrit{\kappa_{\rm c}}
\def\mq{m_{\rm q}}
\def\mqtilde{\widetilde{m}_{\rm q}}
\def\mps{m_{\rm PS}}
\def\mv{m_{\rm V}}
\def\gev{\rm GeV}
\def\mev{\rm MeV}
\def\fm{\rm fm}
\def\phys{{\rm phys}}
\newcommand{\rb}[1]{\raisebox{1.5ex}[-1.5ex]{#1}}
\newcommand{\gtaeq}{\raisebox{-.6ex}{$\stackrel{\textstyle{>}}{\sim}$}}

\begin{document}

\draft        % to get PACS

%\preprint{OUTP--99--XX--P}
%\preprint{Edinburgh Preprint 99/YY\\ OUTP--99--XX--P}
\title{
\begin{flushright}
{\small
OUTP--99--50P \\
Edinburgh 1999/16 \\
DAMTP-1999-138
}
\end{flushright}
Quenched QCD with $O(a)$ improvement: I. The spectrum of light hadrons}
\author{UKQCD Collaboration}
\author{K.C.~Bowler, P.~Boyle\footnote{present address: Department
of Physics \& Astronomy, University of Glasgow, Glasgow G12 8QQ},
J.~Garden, R.D.~Kenway, 
D.G.~Richards\footnote{present address: Jefferson Lab, 12000
Jefferson Avenue, Newport News, VA~23606, and Physics Department, Old
Dominion University, Norfolk, VA~23529}, P.A.~Rowland,
S.M.~Ryan\footnote{present address: School of Mathematics, Trinity
College, Dublin, Ireland}, H.~Simma} 
\address{Department of Physics \& Astronomy, University of Edinburgh,
Edinburgh EH9~3JZ, Scotland}
\author{C.~Michael}
\address{Department of Mathematical
Sciences, University of Liverpool, Liverpool L69~3BX}
\author{H.P.~Shanahan}
\address{DAMTP, University of Cambridge, 21~Silver Street, Cambridge CB3~9EW}
\author{H.~Wittig}
\address{Theoretical Physics, 1~Keble Road, Oxford OX1~3NP}

\date{\today}

\maketitle

\begin{abstract}
We present a comprehensive study of the masses of pseudoscalar and
vector mesons, as well as octet and decuplet baryons computed in
$O(a)$ improved quenched lattice QCD. Results have been obtained using
the non-perturbative definition of the improvement coefficient $\csw$,
and also its estimate in tadpole improved perturbation theory. We
investigate effects of improvement on the incidence of exceptional
configurations, mass splittings and the parameter~$J$. By combining
the results obtained using non-perturbative and tadpole improvement in
a simultaneous continuum extrapolation we can compare our spectral
data to experiment. We confirm earlier findings by the CP-PACS
Collaboration that the quenched light hadron spectrum agrees with
experiment at the 10\,\% level.
\end{abstract}

% Lattice gauge theory, General properties of QCD

\narrowtext

\section{Introduction}

Despite recent efforts in simulating lattice QCD with dynamical
quarks~\cite{burkh_lat98,rdk_lat98,mawh_lat99} the quenched
approximation is still widely used. While precision tests of QCD
through numerical simulations with dynamical quarks are not possible
with the present generation of machines, accurate calculations of
experimentally known quantities, such as the light hadron spectrum,
can be performed using the quenched approximation. Recently the
results of such a benchmark calculation using the Wilson fermion
action have been presented by the CP-PACS
Collaboration~\cite{CP-PACS_quen}, superseding a similar study
performed earlier by GF11~\cite{GF11_quen}. Results from a similar
calculation employing staggered fermions were published
in~\cite{MILC_quen}. In ref.~\cite{CP-PACS_quen} it was concluded that
the quenched light hadron spectrum deviates significantly from
experiment by about 10\%.

In order to reach this level of precision one needs to have control over many
systematic effects, in particular lattice artefacts. In
refs.~\cite{CP-PACS_quen,GF11_quen} extrapolations to the continuum limit were
performed, thus eliminating the dependence on the lattice
spacing~$a$. However, since the leading cutoff effects for Wilson fermions are
linear in~$a$, it is desirable to corroborate these findings and extend the
analysis to weak hadronic matrix elements by performing a similar study using
an improved action.

To leading order in $a$ the Symanzik improvement programme amounts to adding
the well-known Sheikholeslami-Wohlert term to the fermionic Wilson action
\cite{SW85}
\begin{equation}
  \delta S = -\csw\frac{i\kappa}{2}\sum_{x,\mu,\nu}\psibar(x)
  \sigma_{\mu\nu}F_{\mu\nu}(x) \psi(x).
\end{equation}
Provided that $\csw$ is chosen appropriately, spectral quantities such
as hadron masses approach the continuum limit with a rate proportional
to $a^2$. Non-perturbative determinations of $\csw$ have been
performed in the quenched approximation~\cite{AlphaIII,SCRI_imp} and
for $n_f=2$ flavours of dynamical
quarks~\cite{JansenSommer_98}. Estimates of $\csw$ in tadpole improved
perturbation theory \cite{lepenzie93} are also widely used. Results
for quantities in the light hadron sector using one or the other of
the two methods have appeared
recently~\cite{UKQCD_tad57,QCDSF,SCRI_tad,RomeII,SCRI_imp,RomeI,ALPHA99,ALPHA_UKQCD}.

In this paper we present results for the quenched light hadron spectrum in the
continuum limit, using data computed for both non-perturbative and tadpole
improved definitions of $\csw$ at several values of the lattice spacing. By
combining the two datasets and performing a simultaneous continuum
extrapolation, we obtain an independent check of the results reported in
\cite{CP-PACS_quen,GF11_quen}, using a different discretisation of the
theory. Here we concentrate on the light hadron spectrum. Our results for weak 
matrix elements such as decay constants will be published
elsewhere.

The main conclusion of this work is that the previously observed
agreement of the quenched light hadron spectrum with experiment at the
level of 10\,\%, is confirmed. Furthermore, we present qualitative
and quantitative analyses of the effects of $O(a)$ improvement on mass
splittings, the parameter~$J$, the quark mass dependence of hadrons
and the approach to the continuum limit. In many ways this work is a
continuation of a previous paper~\cite{UKQCD_tad57}.

The outline of the paper is as follows: in sect.~\ref{sec_simul} we
present the details of our simulations, including the definition of
improvement coefficients and our numerical
procedures. Sections~\ref{sec_mesons} and~\ref{sec_baryons} contain
discussions of the ``raw'' results in the mesonic and baryonic
sectors, respectively. The quark mass dependence of hadron masses is
discussed in sect.~\ref{sec_quarkmassdep}. In sect.~\ref{sec_contlim}
we present our results extrapolated to the continuum limit. Detailed
comparisons of our results and conclusions are presented in
sect.~\ref{sec_concl}.

\section{Details of the simulation}\label{sec_simul}
\subsection{Improvement coefficients and simulation parameters}

We have generated gauge field configurations using the Wilson
plaquette action at three values of $\beta=6/g_0^2$, namely
$\beta=5.7, 6.0$ and~$6.2$. We used the same hybrid over-relaxed
algorithm described in \cite{light93}. For the fermions we have used
the $O(a)$ improved Wilson action defined by
\begin{eqnarray}
  S_F^{\rm impr}[U,\psibar,\psi]& =& S_F^W[U,\psibar,\psi] \nonumber\\
  & -& \csw\frac{i\kappa}{2} \sum_{x,\mu,\nu}
   \psibar(x)\sigma_{\mu\nu}F_{\mu\nu}(x)\psi(x),
\end{eqnarray}
where $S_F^W$ is the standard Wilson action and $F_{\mu\nu}$ is a lattice
definition of the field strength tensor. The improvement coefficient $\csw$
has been calculated to one loop in perturbation theory \cite{Wohlert,AlphaII}
\begin{equation}
  \csw = 1+0.267\,g_0^2+O(g_0^4).
\end{equation}
It has also been determined non-perturbatively for $\beta\geq6.0$ in
ref.~\cite{AlphaIII} and for $\beta\geq5.7$ in \cite{SCRI_imp}.

We have computed quark propagators at $\beta=6.0$ and 6.2, using the
non-perturbative determination of $\csw$ from \cite{AlphaIII}
\begin{equation}
  c_{\rm sw}^{\rm np} =
  \frac{1-0.656g_0^2-0.512g_0^4-0.054g_0^6}{1-0.922g_0^2}, \quad
  \beta\geq6.0. 
\end{equation}
Furthermore we have used tadpole improved tree-level estimates for $\csw$,
\begin{equation}
  c_{\rm sw}^{\rm tad} = u_0^{-3},\quad  u_0^4=\frac{1}{3}\left\langle 
  {\rm Re\,Tr}\,U_P\right\rangle,
\end{equation}
in order to calculate quark propagators at $\beta=5.7,\,6.0$ and
6.2. In the following we shall refer to the datasets computed using
either $c_{\rm sw}^{\rm np}$ or $c_{\rm sw}^{\rm tad}$ as NP and TAD,
respectively. 

Our values for the hopping parameter $\kappa$ were chosen such that
they straddle the region of the strange quark mass. The simulation
parameters for each dataset are compiled in Table~\ref{tab_simpar},
which also contains the estimates of the spatial extensions for each
lattice in physical units. Exceptional configurations which were
encountered at $\beta=6.0$ for $c_{\rm sw}^{\rm np}$ have been removed
from the statistical ensemble. The incidence of those configurations
is examined in more detail in subsect.~\ref{sec_excep}.

In ref. \cite{AlphaI} it was argued that the bare parameters have to be
rescaled in the $O(a)$ improved theory, so that spectral quantities approach
the continuum limit with a rate proportional to $a^2$.
\footnote{The rescaling is required if a mass independent renormalisation
scheme is adopted in which all renormalisation conditions are imposed at zero
quark mass. In order for such a scheme to be compatible with $O(a)$
improvement the renormalisation of the bare parameters cannot be avoided.}
In the quenched approximation the rescaling needs to be performed only for the 
bare (subtracted) quark mass 
\begin{equation}
\mq = \frac{1}{2a}\left(\frac{1}{\kappa}-\frac{1}{\kcrit}\right),
\end{equation}
where $\kcrit$ is the critical value of the hopping parameter. The rescaled
quark mass $\mqtilde$ is defined by
\begin{equation}
  \mqtilde = \mq\left(1+\bm\,a\mq\right),
\label{eq_mqtildedef}
\end{equation}
and the improvement coefficient $\bm$ has been computed in one-loop
perturbation theory as \cite{AlphaV}
\begin{equation}
  \bm = -\frac{1}{2}-0.0962g_0^2+O(g_0^4).
\label{eq_bmdef}
\end{equation}
So far $\bm$ has been determined non-perturbatively only at
$\beta=6.2$~\cite{GiuPet97}. We have thus used the perturbative
estimate in eq.~(\ref{eq_bmdef}), evaluated with a ``boosted''
coupling $g^2=g_0^2/u_0^4$ unless stated otherwise.  In practice we
found that the details in the evaluation of $\bm$ (e.g. bare versus
boosted perturbation theory) have little influence on our results.

\subsection{Hadron correlators and fitting procedure}

Our quark propagators were calculated using both local and smeared
sources and sinks. The smearing was performed using either the
``fuzzing'' technique described in ref. \cite{fuzzing} or the Jacobi
smearing algorithm of ref. \cite{smearing}. Both smearing procedures
are gauge invariant. They also have a number of parameters, which can
be tuned in order to optimise the projection on a given hadronic
state. For Jacobi smearing the projection properties are controlled by
the parameter $\kappa_S$, which appears in the kernel of the smearing
operator, and the number of iterations,
$N_{\rm{jac}}$~\cite{smearing}. Based on our experience we always
chose~$\kappa_S=0.25$ and used $N_{\rm{jac}}$ to control the smearing
radius. 

The fuzzing algorithm for hadronic correlators has three tunable
parameters, denoted by~$c,\,N_{\rm fz}$ and~$r$. The parameter~$c$ is
the so-called ``link-staple mixing ratio'', which appears in the
construction of fuzzed spatial links (at fuzzing level~$n$) according
to~\cite{albanese87}
\begin{equation}
  U_j^{(n)}(x) = {\cal P}\left\{ cU_j^{(n-1)}(x)
  +\sum_{k=\pm1,k\not=j}
  U_k^{(n-1)}(x)\,U_j^{(n-1)}(x+\hat{k})\,
  U_k^{(n-1)\dagger}(x+\hat{j}) \right\},
\end{equation}
where ${\cal P}$ denotes the projection back into the group manifold
of SU(3). The maximum number of fuzzing levels is given by
$N_{\rm{fz}}$. Throughout this work we have used $c=2$ and
$N_{\rm{fz}}=5$. The size of the fuzzed source (sink) is then
determined by~$r$, which is simply the length of the straight path of
fuzzed links emanating from the origin into all (positive and negative)
spatial directions.

An extensive investigation into the optimal smearing parameters, using
the projection on both mesonic and hadronic states, was performed at
$\beta=6.0$ on $16^3\cdot48$ for
$c_{\rm{sw}}^{\rm{np}}$~\cite{par_thesis}. It was found that
$N_{\rm{jac}}=16$ turned out to be a compromise between good
projection properties and acceptable noise levels in all types of
correlators. Similarly, the optimal radius for fuzzed sources was
determined to be~$r=6$. For different $\beta$-values the radius~$r$
was scaled with the lattice spacing. The type of smearing and the
corresponding values of~$N_{\rm{jac}}$ or~$r$ are listed in
Table~\ref{tab_simpar} for all datasets.

Quark propagators computed using smeared or local sources/sinks were
combined into hadron correlators. We always use the generic notation
'S' to denote correlators which have been smeared, regardless of
whether fuzzing or Jacobi smearing was used to smear the sources
and/or sinks. By 'L' we denote unsmeared (``local'') sources and
sinks. For instance, meson and baryon correlators which have been
smeared at the source but not at the sink are both labelled 'SL' in
this notation. The generalisation to other combinations of source and
sink smearing is obvious.

We have computed meson and baryon correlators for degenerate and
non-degenerate combinations of quark masses. Meson correlators in the
pseudoscalar and vector channels were analysed, as well as spin-1/2 (octet)
and spin-3/2 (decuplet) baryons.

Our meson masses were extracted by performing correlated, simultaneous
fits to the (LL,\,SS) or (LL,\,SL) combination of correlators. In most
cases we used a double-cosh formula to fit the ground state and the
first excitation, requiring the masses in the fit formulae to coincide
for both the LL and SS (or SL) correlators. At $\beta=6.0$ on
$32^3\cdot64$ the double-cosh fits turned out to be unstable, so that
we resorted to single-cosh fits to either the SS or SL correlator. For
baryons we followed the same strategy, using double-exponential fits,
and, at $\beta=6.0$, $32^3\cdot64$, a single exponential.

All fitting intervals have been determined by performing a ``sliding
window'' analysis, in which we first selected the maximum timeslice,
$t_{\rm max}$, of the fitting interval (usually $t_{\rm
max}\;\lesssim\;T/2$) and then pushed $t_{\rm min}$ to its lowest
value which was compatible with the requirements of low $\chi^2/\rm
dof$ and overall stability of the fitted masses.

All statistical errors have been estimated using the bootstrap method with
1000 bootstrap samples. More details about our implementation of the method
can be found in \cite{strange}.

\subsection{Exceptional configurations}
\label{sec_excep}

It has been noted that calculations of fermionic quantities
occasionally suffer from abnormally large fluctuations, in particular
for small quark masses~\cite{exceptional,AlphaIII}. These fluctuations
have been linked to exceptionally small eigenvalues of the Dirac
operator, and the gauge configurations on which they occur are
usually called ``exceptional configurations''. The fraction of such
configurations in the total statistical ensemble increases for smaller
quark mass $\mq$ and/or larger values of $g_0^2$, $c_{\rm sw}$ and the
lattice volume~\cite{AlphaIII}.

In our simulations we have encountered exceptional configurations at
$\beta=6.0$, but not at the other two $\beta$-values. In order to
compare their incidence for $c_{\rm sw}^{\rm tad}$ and~$c_{\rm
sw}^{\rm np}$ we have analysed distributions of observables for
$\beta=6.0$ on $16^3\cdot48$, using the smallest quark mass in the TAD
and NP datasets. The chosen observable was the unsmeared (i.e. LL)
pseudoscalar correlator at $t=T/2$.

To this end we have determined the median $x_m$ and the values
denoting the upper ($x_u$) and lower ($x_l$) ends of the central
68\,\%. As a measure for the width one can define the ratio
\begin{equation}
  \frac{\Delta{x}_u}{x_m},\quad \Delta{x}_u = x_u-x_m.
\end{equation}
The distributions are quite similar for the TAD and NP
datasets. First, their width is comparable, since
${\Delta{x}_u}/{x_m}\approx0.65$ in both cases. Second, both
distributions extend smoothly out to about $x_m+9\Delta{x}_u$.

There are, however, differences in the tails of the distributions,
i.e. the number of values encountered far beyond
$x_m+9\Delta{x}_u$. In the TAD dataset only one configuration is
encountered, which produces a value at roughly $37\Delta{x}_u$ above
the median, whereas in the NP dataset there are three such
configurations with values for the pseudoscalar correlator at
44\,$\Delta{x}_u$, 65\,$\Delta{x}_u$ and~360\,$\Delta{x}_u$ above
$x_m$.

We draw two conclusions from this analysis. The fact that the width is
comparable (i.e. the value of $\Delta{x}_u/x_m$) suggests that the
typical statistical fluctuations do not increase significantly as
$c_{\rm sw}$ is increased from its tadpole improved perturbative
estimate to the non-perturbative value. Second, we have confirmed the
increase of the fraction of exceptional configurations (i.e. those
configurations for which the observable shoots up to values which are
orders of magnitude above the normal level of fluctuations) for larger
$c_{\rm sw}$. The presence of a zero eigenvalue of the Dirac operator
at a nearby $\kappa$-value has also been verified for such
configurations~\cite{simsmi_98}. 

We did not make attempts to treat exceptional configurations using,
for instance, the methods described
in~\cite{exceptional_treat}. Instead we have chosen to eliminate them
from our statistical ensemble. That is, at $\beta=6.0$, $16^3\cdot48$
we have removed the two configurations which produced the most extreme
values in the distribution of the unsmeared pion propagator for
$c_{\rm sw}^{\rm np}$, and on which the inversion of the Dirac
operator did not converge for some of its components. The latter also
occurred on another configuration if a fuzzed source was used, and
that configuration was subsequently removed as well. The total number
of exceptional configurations which were removed for a particular
dataset are shown in brackets in Table~\ref{tab_simpar}. Note that no
configurations were eliminated from the TAD datasets.

For our range of $\beta$-values and the corresponding values
of~$c_{\rm sw}$ the incidence of exceptional configurations is still
relatively small. Their fraction in the NP dataset amounts to less
than 1\% on $16^3\cdot48$ (3\% on $32^3\cdot64$), and after the
analysis presented here we do not expect serious distortions of the
statistical ensembles due to their removal.

\section{Results for pseudoscalar and vector masses}
\label{sec_mesons}

In Tables\,\ref{tab_ps_NP}--\ref{tab_vec_TAD} we present our ``raw''
results for meson masses in the pseudoscalar and vector channels,
which were obtained from the fits described in the previous section.
The fit ranges were determined independently for the (LL,\,SL) and
(LL,\,SS) combinations of correlators. Both combinations gave
consistent results, and in general we quote the result from the fit
which gave the best value of $\chi^2/{\rm dof}$.

In the pseudoscalar channel the fits were very stable under
variations of the fitting interval. By contrast, the fits in the
vector channel could in some cases differ by up to one standard
deviation if a different fitting interval was selected. We estimate
the systematic error in the mass of the vector meson arising from
choosing alternative fitting ranges to be at most as large as the
statistical error. This systematic error has not been included in
the tables.

\subsection{Finite volume effects}

Based on our results for the NP dataset obtained at $\beta=6.0$ on
$16^3\cdot48$ and $32^3\cdot64$, we can make a first estimate of
finite size effects in the mesonic sector. In physical units the
spatial extensions of the two lattices correspond to
$L\approx1.5\,\fm$ and~$3.0\,\fm$, respectively.

In the pseudoscalar channel we find evidence for small but significant
finite volume effects. On the smaller lattice the values for $a\mps$
are consistently larger. Furthermore, the effect shows a trend to
increase as the quark mass gets smaller. Both these observations are
consistent with the expected qualitative features of finite-size
effects. The difference in $a\mps$ determined for the two lattice
sizes amounts to 0.6\% at the largest and 1.5\% at the smallest quark
mass. At all values of $\kappa$ the deviation is a two-$\sigma$
effect.

By contrast, no statistically significant finite size effects are
observed in the vector channel. In fact, the values for $a\mv$ are
slightly higher on the larger lattice. This might be attributed to the
fact that no estimate for contributions from excited states was
available for all hadron masses computed on $32^3\cdot64$, since only
single-cosh fits could be performed. Indeed, one of the caveats in the
analysis of finite volume effects in both channels is the fact that
the data for the two lattice sizes shown in Tables~\ref{tab_ps_NP}
and~\ref{tab_vec_NP} have not been obtained using the same fitting
procedure. We have, therefore, repeated the analysis for the smaller
volume, by performing single-cosh fits for appropriately chosen
intervals. The results are consistent with those shown in the tables.
However, in the pseudoscalar channel it is also possible to choose
small fitting intervals close to $t_{\rm max}$ such that the
single-cosh fits on $L\approx1.5\,\fm$ produce smaller values with
larger errors, which are both compatible with the results in the
tables and also with those obtained on the larger volume. We conclude
that the finite size effects observed in the pseudoscalar channel
appear to be genuine, but without further investigations one cannot
rule out entirely that they have a statistical origin. For vector
mesons no significant effects are observed.

\subsection{Vector-pseudoscalar mass splitting}

It has been known for some time that lattice results obtained in the 
quenched approximation fail to reproduce the experimental fact that
the vector-pseudoscalar hyperfine splitting is constant over a wide
range of quark masses, i.e. $\mv^2-\mps^2\approx
0.55\,\gev^2$. Indeed, lattice estimates for this quantity are in
general much lower when unimproved Wilson fermions are employed. 

\begin{figure}[tp]
\vspace{1.0cm}
\setlength{\epsfxsize}{16cm}\epsfbox[18 350 554 769]{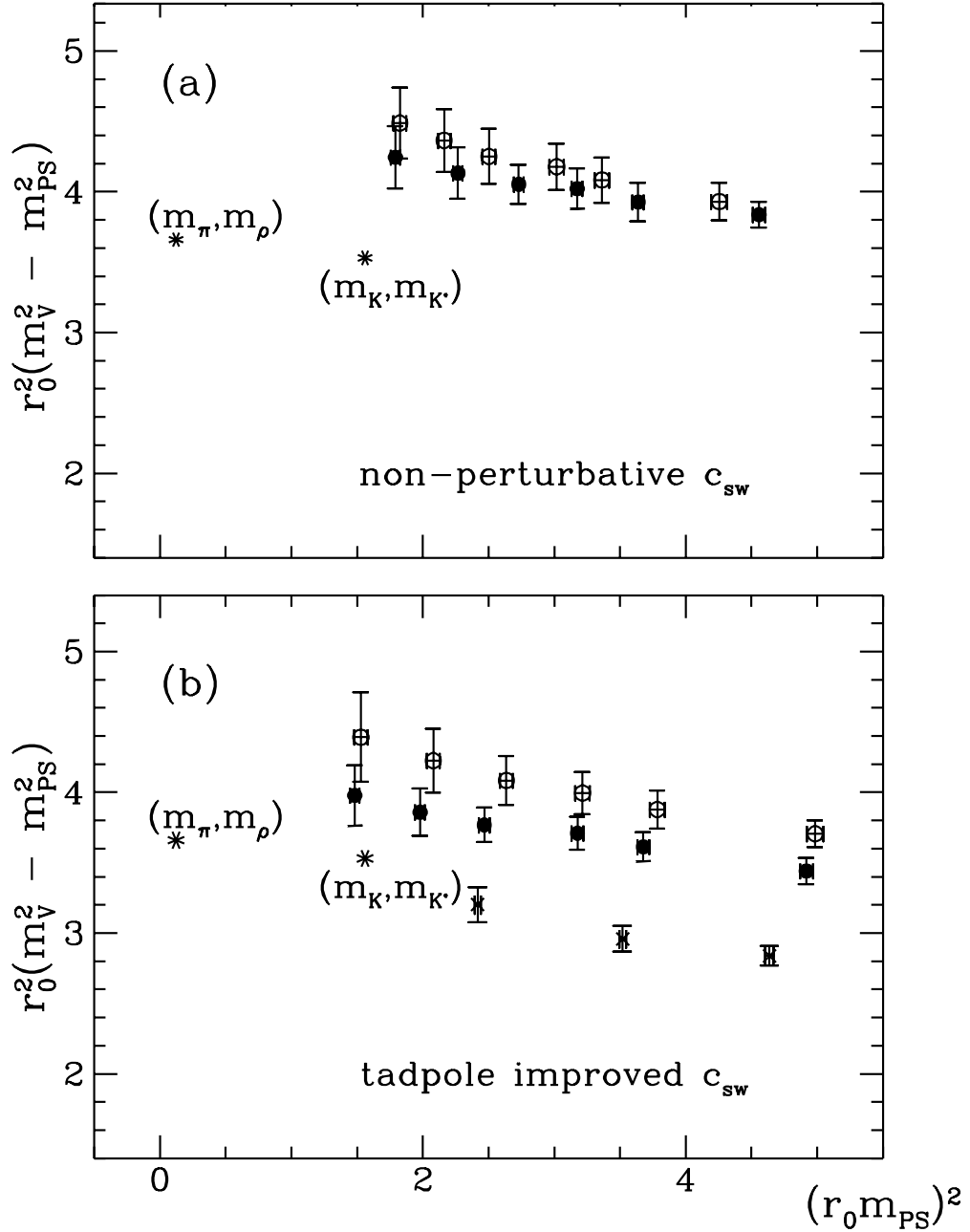}
\vspace{6.5cm}
\caption{The vector-pseudoscalar hyperfine splittings for (a): the NP
and (b): the TAD datasets. Open squares, full circles and crosses
denote the data at $\beta=6.2,\,6.0$ and~5.7, respectively. The
experimental points are represented by the asterisks.}
\label{fig_splittings}
\end{figure}

In order to study the effect of $O(a)$ improvement on the hyperfine
splitting we have plotted our results for the NP and TAD datasets in
Fig.\,\ref{fig_splittings}\,(a) and\,(b), respectively. In order to
display the influence of finite lattice spacing, we have expressed
our results in units of the hadronic radius $r_0$\,\cite{sommer_r0},
using its lattice determination in ref.\,\cite{alpha_r0}. The figure
demonstrates that with improvement (either non-perturbative or mean
field) the hyperfine splittings show much weaker variation over the
studied range of quark masses, compared to the unimproved case (see,
e.g. ref. \cite{light93}). However, the small slope in the data for
$r_0^2(\mv^2-\mps^2)$ as a function of $(r_0\mps)^2$ suggests that
the experimentally observed {\it very} weak dependence on the quark
mass is not reproduced by the lattice data.
%Hence, despite the fact
%that lattice results are close to the experimental values in the
%region of the strange quark mass, this is unlikely to be the case
%for heavy quarks.

By comparing the NP and TAD datasets it appears that the dependence
on the lattice spacing is somewhat smaller for non-perturbative
$\csw$. Of course this needs to be corroborated in a real scaling
analysis at a fixed value of $(r_0\mps)^2$.

At first sight it may seem surprising that the lattice results for the
hyperfine splitting overestimate the experimentally observed
values. However, lattice results for the splittings in the quenched
approximation in physical units depend strongly on the choice of
scale. Indeed, if the scale is set using $m_K$ the lattice values are
much closer to experiment \cite{how_lat97,par_thesis}. However, the
main focus of this discussion is the analysis of the dependence on the
quark mass and the lattice spacing.

\subsection{The parameter $J$}

The parameter $J$ was introduced \cite{UKQCD_J} as a means to detect
deviations between the quenched approximation and the observed
hadron spectrum without relying on chiral extrapolations. It is
defined through
\begin{equation}
 J = \mv\frac{d\mv}{d\mps^2},\quad \mv/\mps=m_{K^*}/m_K,
\end{equation}
and is thus related closely to the slope of the vector-pseudoscalar
splitting discussed above. Its phenomenological value has been
determined from the experimentally measured masses as $J=0.48(2)$.

\begin{figure}[tp]
\vspace{1.0cm}
\setlength{\epsfxsize}{16cm}\epsfbox[18 180 554 589]{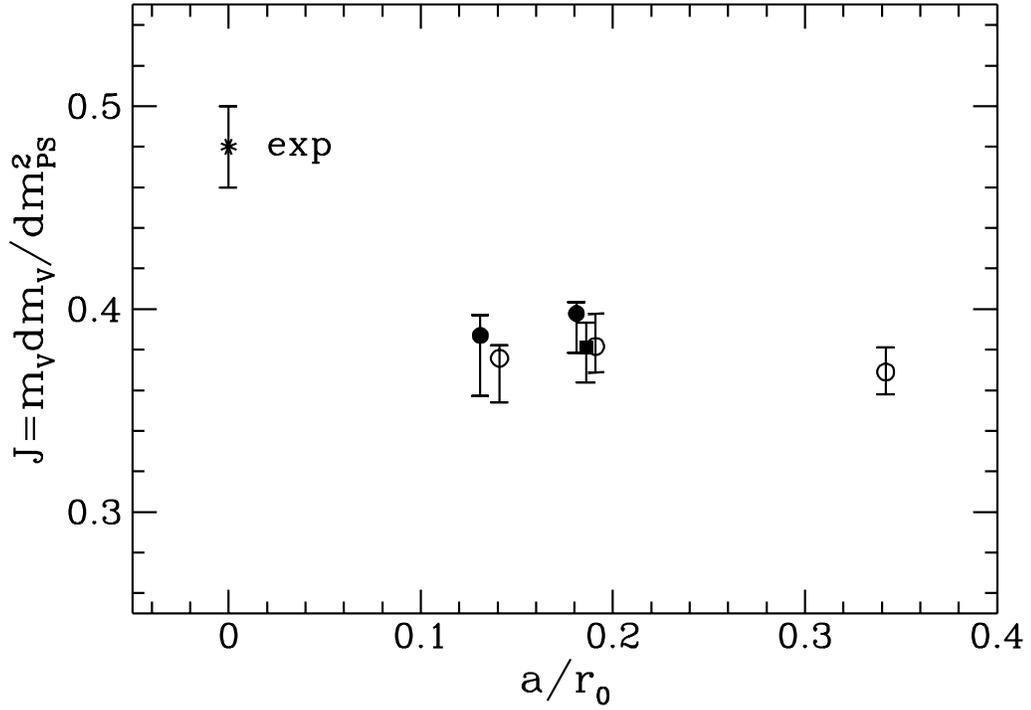}
\vspace{-1.5cm}
\caption{The parameter $J$ plotted versus the lattice spacing in
units of $r_0$. Filled (open) symbols denote the data using
non-perturbative (tadpole improved)
$\csw$. The filled square denotes the data point at $\beta=6.0$ on
the larger volume of $32^3\cdot64$.}
\label{fig_J_all}
\end{figure}

In Fig.\,\ref{fig_J_all} we plot our results for all our datasets as 
a function of the lattice spacing. Our values confirm previous
observations that $J$ is underestimated in the quenched
approximation. In fact, one finds that the low values for $J$ have
little to do with lattice artefacts, since there is no sign of the
data approaching the phenomenological value for $J$ in the continuum 
limit. We conclude that low lattice estimates for $J$ appear to be an 
intrinsic feature of the quenched approximation.

%\subsection{Current quark masses}
%
%\P {\it Optional. We could discuss $\kcrit$ extracted from the PCAC
%mass, but I'm less in favour of publishing our values for the
%current quark mass at this stage, since we might include them in the
%second paper on matrix elements (decay constants and quark
%masses).}\P

\section{Results for octet and decuplet baryons}
\label{sec_baryons}

Our results for masses of octet and decuplet baryons are shown in
Tables~\ref{tab_bardeg_NP}--\ref{tab_barnondeg_TAD_62}. They have been
obtained by performing double-exponential fits to the (LL,SS) or
(LL,SL) combination of baryon correlation functions, except on
$32^3\cdot64$ for the NP dataset at $\beta=6.0$, where -- in analogy
to the fits of the large-volume data in the mesonic sector -- only
single exponential fits to smeared correlators were considered.

Tables~\ref{tab_bardeg_NP} and~\ref{tab_bardeg_TAD} contain the
results for baryon masses in the nucleon $(J^P={\frac{1}{2}}^{+})$
and $\Delta$ $(J^P={\frac{3}{2}}^{+})$ channels for degenerate
combinations of quark masses, together with the fitting ranges and
the values of $\chi^2/{\rm dof}$. These results can be extrapolated
to the physical values of the quark masses in order to determine the
masses of the nucleon, the $\Delta(1232)$ and the $\Omega(1672)$, as 
described in sect.~\ref{sec_quarkmassdep}.

\subsection{Baryons for non-degenerate quark masses}

In order to compute the masses of the physical $\Lambda, \Sigma$ and
$\Xi$ states one has to consider baryon correlators for non-degenerate
combinations of quark masses. In the octet sector one has to
distinguish between ``$\Sigma$-like'' and ``$\Lambda$-like''
correlators. Using the generic notation $u,d,s$ to denote quark
flavours, we note that $\Sigma$-like states are symmetric in the light
flavours $u,\,d$, whereas $\Lambda$-like states are antisymmetric. On
the lattice the corresponding correlators are obtained by performing
the appropriate contractions
\cite{LabShar,BhatGupKilSha95,par_thesis}. The $J^P={\frac{1}{2}}^{+}$
states of the $\Sigma$ and $\Lambda$ are then obtained from the
correlation functions by averaging the 11 and 22 spinor indices.

The correlators for decuplet baryons, which are symmetric in all
flavours $u,d,s$ are simpler to construct. They are obtained in
terms of the interpolating operator
\begin{equation}
   D_{\mu;ijk} = \epsilon_{abc}\left(\psi^a_i C\gamma_\mu
   \psi^b_j\right) \psi^c_k,
\end{equation}
where $i,j,k$ denote the quark flavour, $a,b,c$ are colour indices,
and $C$ is the charge conjugation matrix. Correlation functions for
decuplet baryons are constructed from the correlation of
$D_{\mu;ijk}$ by projecting out the spin-$\frac{3}{2}$ component.

In Tables~\ref{tab_barnondeg_NP_60small}--\ref{tab_barnondeg_TAD_62}
we list the results for octet ($\Sigma,\Lambda$-like) and decuplet
($\Delta$-like) baryons for non-degenerate combinations of quark
masses. The fitting intervals, which are not shown, are mostly
identical to those chosen for the corresponding channels in the
degenerate case (Tables~\ref{tab_bardeg_NP}
and~\ref{tab_bardeg_TAD}). With our statistical accuracy we are not
able to distinguish between $\Sigma$- and $\Lambda$-like states; the
different symmetry properties in the quark flavours corresponding to
the hopping parameters $\kappa_1$ and $\kappa_2$ do not manifest
themselves in statistically significant mass differences.

By extrapolating or interpolating the data in $\kappa_1, \kappa_2,
\kappa_3$ to the hopping parameters corresponding to the physical
quark masses one obtains the masses of the $\Lambda, \Sigma, \Sigma^*,
\Xi, \Xi^*$ and~$\Omega$. Note that non-degenerate combinations of
quark masses have not been computed at $\beta=5.7$.

\subsection{Finite volume effects in the baryonic sector}

The issue of finite-volume effects is of special importance in the
baryonic sector where these effects are expected to be more severe
than for mesons. With our data we can assess the influence of
finite-volume effects by comparing our results computed at
$\beta=6.0$ on either $16^3\cdot48$ or $32^3\cdot64$ using the
non-perturbative value of $\csw$. 

The numbers in Tables~\ref{tab_bardeg_NP},
\ref{tab_barnondeg_NP_60small} and~\ref{tab_barnondeg_NP_60large}
suggest that on the large volume the mass estimates for both octet
and decuplet baryons are slightly smaller. For octet baryons this
decrease amounts to about 1.4\% at the largest and 2.7\% at the
smallest quark masses. The effect is roughly twice as large as for
the pseudoscalar mesons discussed earlier. Although finite-size
effects for octet baryons are not significant at our level of
statistical accuracy, this does not necessarily indicate that those
effects are absent.

For decuplet baryons the finite-volume effects are more pronounced;
they vary between~2.4 and~5.5\%, again increasing towards smaller
quark masses. Here the discrepancy between the results on the small
and large volumes amounts to about 1.5 standard deviations. Thus we
cannot exclude finite-size effects in our baryon data at a level of up
to 2.5\% for octet and 5.5\% for decuplet baryons.

\section{Quark mass dependence}
\label{sec_quarkmassdep}

In this section we discuss the dependence of mesons and baryons on the
quark mass. Usually this dependence is modelled using the results of
chiral perturbation theory at lowest order. It is then quite a
delicate problem to decide whether higher orders in the chiral
expansion have to be included. Furthermore, additional care must be
taken in the quenched approximation, where one expects deviations from
the leading behaviour for very small quark masses (i.e. close to the
chiral limit), due to the appearance of quenched chiral
logarithms~\cite{sharpe1,BerGol,LabShar,booth}. We will first motivate
the functional forms for the quark mass dependence used in this paper,
determine the critical hopping parameter and then present our results
for hadron masses extrapolated or interpolated to the physical quark
masses. 

\subsection{Fit ansatz and the critical hopping parameter}

Usually the critical value of the hopping parameter,
$\kcrit$, is determined at the point where the mass 
of the pseudoscalar meson vanishes, $\mps=0$. The simplest ansatz
for the quark mass dependence of $\mps$, which is consistent with
$O(a)$ improvement, is
\begin{equation}
\mps^2 = B(\widetilde{m}_{\rm q,1}+\widetilde{m}_{\rm q,2}),
\label{eq_mps_chiral}
\end{equation}
where $\widetilde{m}_{{\rm q},i}, i=1,2$ denotes the rescaled, bare
quark mass defined in eq.\,(\ref{eq_mqtildedef}).

Assuming that the ansatz in eq.\,(\ref{eq_mps_chiral}) is justified
(i.e. both higher orders in the quark mass as well as quenched chiral
logarithms are assumed to be absent), we have determined $\kcrit$
using both degenerate and non-degenerate combinations of quark masses,
by inserting the definition of $\mqtilde$ into
eq.\,(\ref{eq_mps_chiral}), which leads to the general fit ansatz
\begin{equation}
   \mps^2 = \alpha +
   \beta\left(\frac{1}{\kappa_1}+\frac{1}{\kappa_2}\right) +
   \gamma\left(\frac{1}{\kappa_1^2}+\frac{1}{\kappa_2^2}\right),
\end{equation}
where the fit parameters $\alpha,\,\beta$ and $\gamma$ are related
to $B,\,\kcrit$ and $\bm$ through
\begin{equation}
  \alpha=\frac{B}{\kcrit}\left(-1+\frac{\bm}{2\kcrit}\right),\;
  \beta=\frac{B}{2}\left(1-\frac{\bm}{\kcrit}\right),\;
  \gamma=\frac{B\bm}{4}.
\end{equation}
As mentioned in sect.~\ref{sec_simul} we have used the tadpole
improved perturbative estimate at one loop for $\bm$. In order to
study the sensitivity of $\kcrit$ on $\bm$ we have also used its
estimate in one-loop perturbation theory in the bare coupling, as well
as its tree-level value, $\bm=-1/2$, and, at $\beta=6.2$ for the NP
dataset, the non-perturbative determination of
ref.~\cite{GiuPet97}. In order to enable direct comparisons with
$\kcrit$-estimates from earlier simulations using tadpole improvement,
we have also computed $\kcrit$ for $\bm=0$ for the TAD dataset. Our
results, which are collected in Table~\ref{tab_kappac}, show little
dependence on the value of $\bm$. The largest deviations are observed
at $\beta=5.7$.  Furthermore, using the non-perturbative value of
$\bm$ at $\beta=6.2$ yields a result which is entirely compatible with
the ones obtained using one-loop perturbative estimates.

We conclude that for our range of quark masses, estimates for $\bm$
based on one-loop perturbation theory are sufficient to obtain stable
results for $\kcrit$ for $\beta\;\gtaeq\;6.0$.

\begin{figure}
\vspace{1.0cm}
\setlength{\epsfxsize}{16cm}\epsfbox[18 350 554 769]{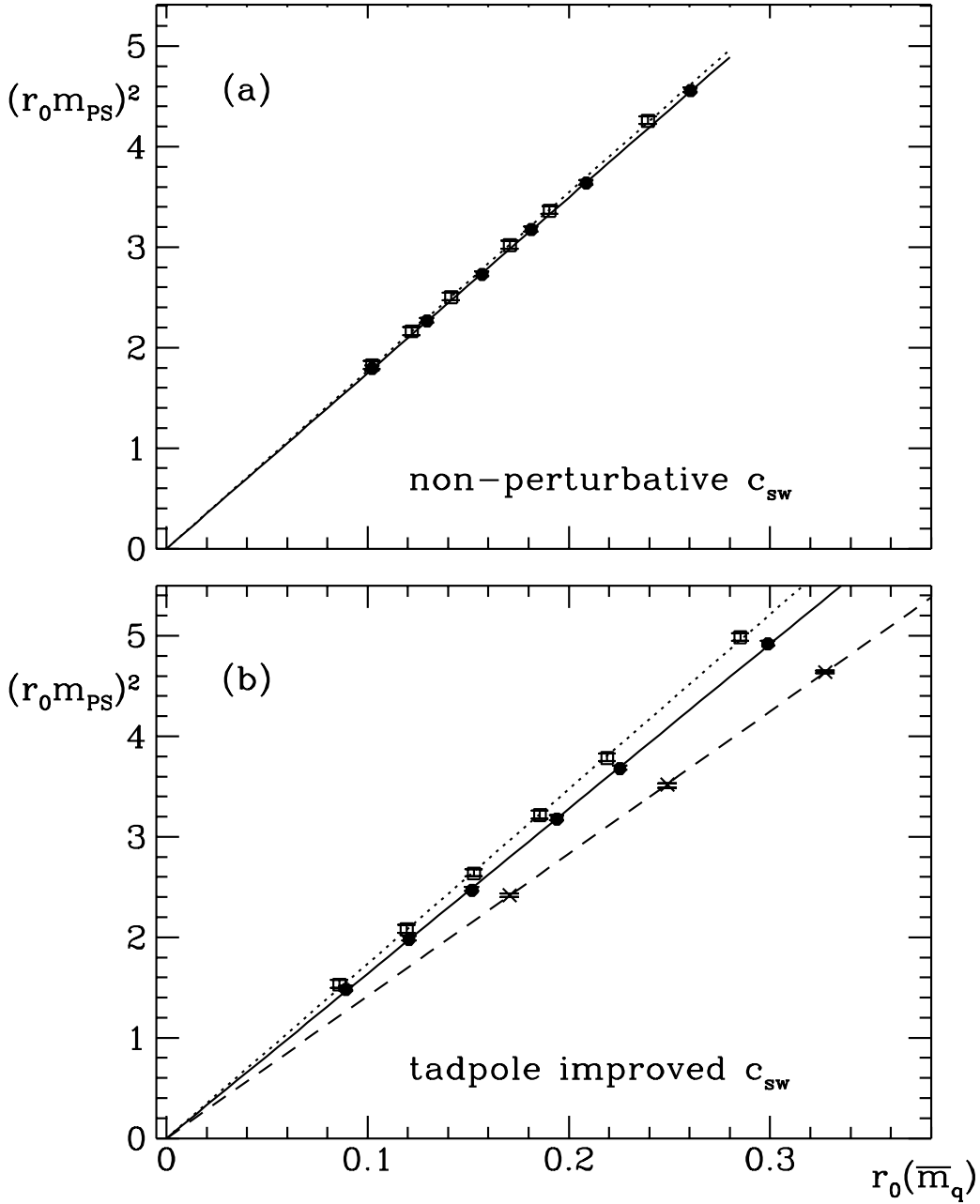}
\vspace{6.5cm}
\caption{The pseudoscalar squared plotted versus the averaged quark
mass, $\overline{m}_q=(\widetilde{m}_{\rm q,1}+\widetilde{m}_{\rm
q,2})/2$ in units of $r_0$ for (a): the NP and (b): the TAD
datasets. Open squares, full circles and crosses denote the data at
$\beta=6.2,\,6.0$ and~5.7, respectively.}
\label{fig_mpssq_vs_mqtilde}
\end{figure}

We can now justify our ansatz eq.\,(\ref{eq_mps_chiral}) by plotting
$\mps^2$ as a function of $(\widetilde{m}_{\rm q,1}+\widetilde{m}_{\rm
q,2})/2$ in units of $r_0$ for both NP and TAD datasets. This is shown
in Fig.~\ref{fig_mpssq_vs_mqtilde} where the lines denote the fits
based on eq.\,(\ref{eq_mps_chiral}). The plots show that no
significant departure from the linear behaviour predicted by lowest
order chiral perturbation theory is observed in the range of quark
masses investigated. Thus, we find no evidence for higher order terms
in the chiral expansion of $\mps^2$, nor do our data support the
presence of quenched chiral logarithms. The latter is most probably
due to the fact that the quark masses used in our simulations are not
light enough.

Furthermore, we wish to point out that a more sophisticated
analysis~\cite{ALPHA_UKQCD} of the quark mass dependence of the data
in Table~\ref{tab_ps_NP} revealed that higher-order terms proportional
to $(\widetilde{m}_{\rm q,1}-\widetilde{m}_{\rm q,2})^2$ contribute
below the 1\% level.

\begin{figure}
\vspace{1.0cm}
\setlength{\epsfxsize}{16cm}\epsfbox[18 350 554 769]{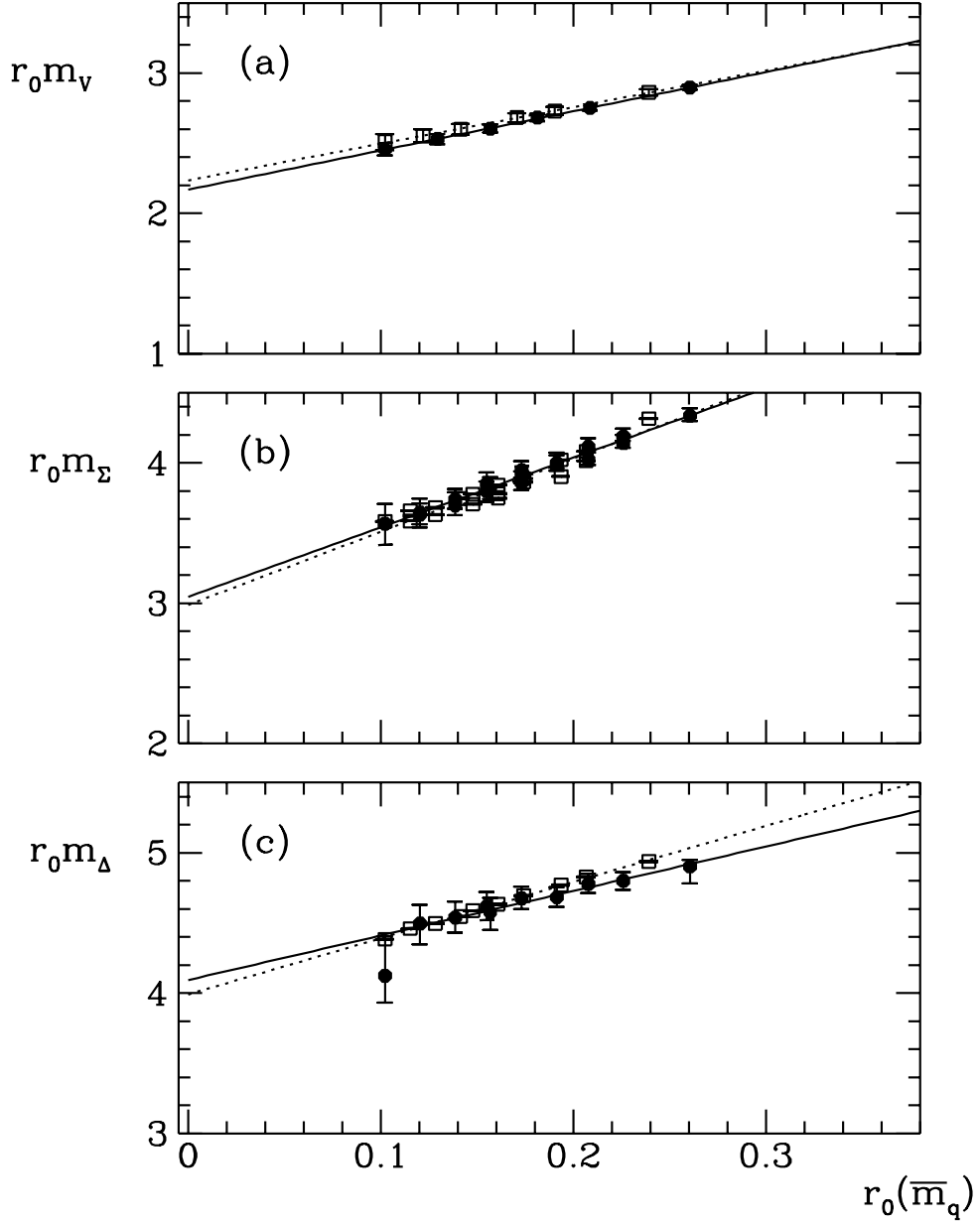}
\vspace{6.5cm}
\caption{Data for (a): vector mesons, (b): $\Sigma$-like baryons and
(c): $\Delta$-like baryons plotted versus the averaged quark mass
$\overline{m}_q=(\widetilde{m}_{\rm q,1}+\widetilde{m}_{\rm
q,2}+\widetilde{m}_{\rm q,3})/3$ in units of $r_0$ for the
non-perturbatively improved (NP) dataset. Open squares and full
circles denote the data at $\beta=6.2$ and~6.0, respectively. The
lines represent the fits to
eqs.~(\protect\ref{eq_mv_chiral}--\protect\ref{eq_mdec_chiral}).}
\label{fig_m_all_vs_mqtilde_np}
\end{figure}

\begin{figure}
\vspace{1.0cm}
\setlength{\epsfxsize}{16cm}\epsfbox[18 350 554 769]{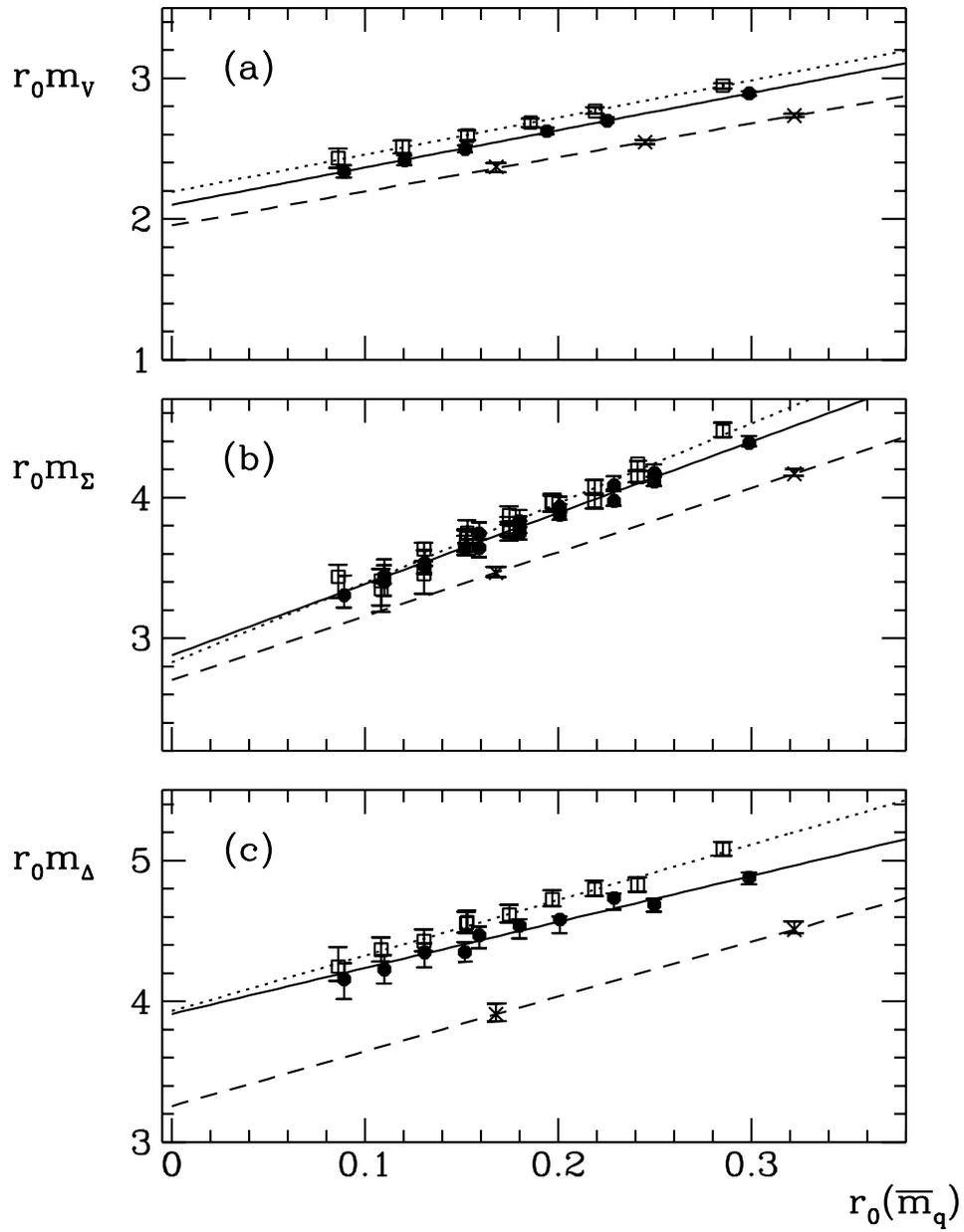}
\vspace{6.5cm}
\caption{Same as Fig.\,\protect\ref{fig_m_all_vs_mqtilde_np} for the
TAD dataset. Open squares, full circles and crosses denote the data at
$\beta=6.2,\,6.0$ and~5.7, respectively.}
\label{fig_m_all_vs_mqtilde_tad}
\end{figure}

Based on our observations in the pseudoscalar channel, which usually
offers the most precise information about the quark mass dependence,
we have used the following functional forms for vector mesons, octet
and decuplet baryons:
\begin{eqnarray}
  \mv & = & A_{\rm V} +C_{\rm V}\left( \widetilde{m}_{\rm q,1}
                              +\widetilde{m}_{\rm q,2} \right)
                              \label{eq_mv_chiral}\\ 
  m_{\rm Oct} & = & A_{\rm O} +C_{\rm O}\left( \widetilde{m}_{\rm q,1}
              +\widetilde{m}_{\rm q,2}+\widetilde{m}_{\rm q,3} \right)
                              \label{eq_moct_chiral}\\ 
  m_{\rm Dec} & = & A_{\rm D} +C_{\rm D}\left( \widetilde{m}_{\rm q,1}
              +\widetilde{m}_{\rm q,2}+\widetilde{m}_{\rm q,3} \right).
                              \label{eq_mdec_chiral}
\end{eqnarray}
The corresponding fits are shown in
Figs.~\ref{fig_m_all_vs_mqtilde_np} and~\ref{fig_m_all_vs_mqtilde_tad}
for the NP and TAD datasets, respectively. As in the case of
pseudoscalar mesons we observe that for our level of precision and
range of quark masses there is no evidence for curvature in the
data. We conclude that
eqs.~(\ref{eq_mps_chiral}--\ref{eq_mdec_chiral}) represent appropriate
fitting functions for our data. A detailed analysis of more
complicated models for the quark mass dependence described
in~\cite{par_thesis} resulted in similar findings.

In a given channel (i.e. pseudoscalar and vector mesons, $\Sigma$-,
$\Lambda$- and $\Delta$-like baryons) we have determined the
parameters $B,\,A_{\rm V},\,C_{\rm V},\ldots$ from uncorrelated,
simultaneous fits to degenerate and non-degenerate combinations of
quark masses. The only exception was the dataset at $\beta=5.7$, for
which only two degenerate combinations of quark masses had been
computed in the baryonic channels. Therefore, the quark mass
dependence at $\beta=5.7$ is not really controlled. Nevertheless, we
have included the results in the following analysis.

\subsection{Hadrons at physical values of the quark masses}

Our task now is to make contact with the physical hadron spectrum by
matching the quark masses in
eqs.~(\ref{eq_mps_chiral}--\ref{eq_mdec_chiral}) to the masses of the
physical $u,d$ and~$s$ quarks. Here we employ the axial Ward
identities, which, for the physical pseudoscalar mesons (in the
continuum theory), read
\begin{eqnarray}
  m_{\pi^\pm}^2 &=& B(m_u+m_d) \\
  m_{K^\pm}^2   &=& B(m_u+m_s),\qquad m_{K^0}^2 = B(m_d+m_s).
\end{eqnarray}
By assuming isospin symmetry, i.e. $m_u=m_d$ one can define the
so-called ``normal'' quark mass $m_n$ through
\begin{equation}
  m_n \equiv \textstyle\frac{1}{2}(m_u+m_d)
\end{equation}
and the isospin-averaged combination of kaon masses as
\begin{equation}
  m_K^2 \equiv \textstyle\frac{1}{2}(m_{K^\pm}^2+m_{K^0}^2),
\end{equation}
such that
\begin{eqnarray}
  m_{\pi^\pm}^2 &=& 2B\,m_n \\
  m_K^2         &=& B\,(m_n+m_s).
\end{eqnarray}
Inserting $m_{K^\pm}^2=493.7\,\mev$ and
$m_{K^0}=497.7\,\mev$~\cite{PDG98} one obtains\footnote{In principle
one also has to compensate for the electromagnetic binding energy of
about $-0.7\,\mev$ in eq.~(\protect\ref{eq_mKphys}) (see
e.g. ref.~\protect\cite{ALPHA_UKQCD}). However, this has not been done
in this paper.}
\begin{equation}
       m_K=495.7\,\mev.
\label{eq_mKphys}
\end{equation}
Using our lattice estimates for the parameter~$B$ we can determine the
combination of (bare) quark masses $\widetilde{m}_s+\widetilde{m}_n$
and the quark mass $\widetilde{m}_n$ through
\begin{eqnarray}
    a(\widetilde{m}_s+\widetilde{m}_n) & = &
    (aQ)^2\frac{(m_K/Q)_\phys^2}{aB} \\
    a\widetilde{m}_n & = & (aQ)^2\frac{(m_\pi/Q)_\phys^2}{2aB},
\end{eqnarray}
where $Q$ is the quantity which sets the lattice scale, and the
subscript ``phys'' denotes that the physical ratio is to be taken. It
is a well-known fact that in the quenched approximation there is an
intrinsic ambiguity associated with the lattice scale. In order to
estimate this ambiguity we have used three different quantities for
$Q$, namely
\begin{eqnarray}
  Q=r_0^{-1}, &\quad& r_0 = 0.5\,\fm,\;r_0^{-1}=395\,\mev, \nonumber\\
  Q=m_{K^*},  &\quad& m_{K^*}=893.9\,\mev, \nonumber\\
  Q=m_N,      &\quad& m_N=938\,\mev.
\end{eqnarray}
Here, the value $m_{K^*}=893.9\,\mev$ is the isospin averaged result,
i.e. $m_{K^*}=\frac{1}{2}(m_{K^{*\pm}}+m_{K^{*0}})$. Lattice data for
$r_0/a$ and its error at all relevant $\beta$-values were taken from
ref.~\cite{alpha_r0}. In particular, we used the interpolating
formula, eq.~(2.18) of~\cite{alpha_r0}.

The following comments apply to our chosen set of lattice scales:
\begin{itemize}
\item On the lattice the mass of the $K^*$ meson is obtained through
      an interpolation of data points, which is intrinsically a safe
      procedure. However, the physical $K^*$ is an unstable hadron
      with a finite width; resonance effects are not controlled in the
      lattice calculation;
\item The nucleon is a stable hadron, but in lattice simulations its
      mass is obtained only by an extrapolation close to the chiral
      limit. In view of the possible presence of quenched chiral
      logarithms this extrapolation is hard to control. Furthermore,
      precise lattice determinations of baryon masses are more
      difficult compared to the mesonic sector, due to larger
      statistical errors and the possibility of relatively large
      finite size effects;
\item The hadronic radius $r_0$ is known accurately for a wide range
      of lattice spacings~\cite{alpha_r0}, but a direct experimental
      measurement is not available. Its phenomenological value of
      $r_0=0.5\,\fm$ is estimated from potential models fitted to
      experimental data.
\end{itemize}

In Table~\ref{tab_kappa_phys} we have collected the results for the
hopping parameters~$\kappa_n$ and~$\kappa_s$, corresponding to the
quark masses $a\widetilde{m}_n$ and $a\widetilde{m}_s$, obtained for
our three different choices of~$Q$ and using the tadpole improved
perturbative estimate for $\bm$ in the definition of $a\mqtilde$.
Thus, in spite of the difficulties associated with extrapolations to
the chiral limit we have chosen to compute $a\widetilde{m}_n$ and to
quote lattice estimates for $m_\rho$, $m_N$ and $m_\Delta$.

The physical vector meson, octet and decuplet baryon masses have been
computed by inserting the appropriate combinations
of~$a\widetilde{m}_n$ and~$a\widetilde{m}_s$, corresponding to the
physical quark content, into
eqs.~(\ref{eq_mv_chiral}--\ref{eq_mdec_chiral}). The results, in units
of either $r_0,\,m_{K^*}$ or~$m_N$ are shown in
Tables~\ref{tab_res_r0}, \ref{tab_res_mkstar} and~\ref{tab_res_mnuc},
respectively. These tables contain the data with $L/a\leq24$ only. For
completeness, the non-perturbatively improved data obtained on
$32^3\cdot64, \beta=6.0$, which have not been used in the continuum
extrapolation, are listed in Table~\ref{tab_all_res_B60LARGE}.

It has been observed that several tests of the quenched approximation
can be performed at intermediate values of the quark mass (e.g. near
the strange quark mass), so that extrapolations to the chiral limit as
a reference point are not required~\cite{UKQCD_J}. Furthermore, it has
been suggested that properties of the effective chiral Lagrangian can
be studied for {\it unphysical\/} quark masses. A convenient reference
point for future lattice studies is then provided by the condition
$(\mps{r_0})^2 = 3.0$, which has already been chosen in
ref.~\cite{ALPHA_UKQCD}. We have interpolated our results in the
vector meson, $\Sigma$ and $\Delta$ channels to that point by defining
a reference quark mass $\widetilde{m}_{\rm ref}$ through
\begin{equation}
  2(aB)\,(a\widetilde{m}_{\rm ref})\,(r_0/a)^2 = 3.0
\end{equation}
and inserting its value into
eqs.~(\ref{eq_mv_chiral}--\ref{eq_mdec_chiral}). The results are shown
in Table~\ref{tab_res_mref}.

Another reference point which does not require extrapolations to the
chiral limit is defined at the point where
\begin{equation}
        m_{\rm PS}/m_{\rm V} = 0.7.
\label{eq_PSV07}
\end{equation}
Results for the vector meson and nucleon masses at the reference point
defined by eq.~(\ref{eq_PSV07}) have been quoted
in~\cite{SCRI_tad,SCRI_imp}. Our estimates in the vector, $\Sigma$ and
$\Delta$ channels are included in Table~\ref{tab_res_mref}. By
comparing the results in Tables~\ref{tab_res_mref}
and~\ref{tab_res_r0} one observes that both reference points
correspond to the case of degenerate light quarks with masses around
that of the strange quark.

\section{The continuum limit}
\label{sec_contlim}

We are now in a position to discuss the extrapolation of our results
to the continuum limit. This will finally enable us to make a direct
comparison with experimental data and the results of
ref.~\cite{CP-PACS_quen}, obtained using the unimproved Wilson
action. 

For hadron masses computed using the non-perturbative determination of
$\csw$ the leading cutoff effects are expected to be of
order~$a^2$. Detailed scaling studies have confirmed that the approach
to the continuum limit for spectral
quantities~\cite{SCRI_imp,jochen99} and matrix
elements~\cite{jochen99} is indeed consistent with such a leading
order. By contrast, it is expected that small terms of order~$a$
cannot be excluded when tadpole improvement is used~\cite{SCRI_tad}.

From our list of simulation parameters in Table~\ref{tab_simpar} it is
clear that separate continuum extrapolations of the results for the NP
and TAD datasets are not feasible. We have therefore chosen to perform
simultaneous extrapolations by assuming leading lattice artefacts of
order~$a^2$ for the NP and artefacts of both order~$a$ and~$a^2$ for
the TAD dataset. The ansatz for the continuum extrapolation of a
generic hadron mass~$M$ in units of $r_0$ then reads
\begin{equation}
  r_0\,M = \left\{\begin{array}{l}
                  r_0\,M\big|_{a=0} +B^{\rm NP}(a/r_0)^2 \\
                  r_0\,M\big|_{a=0} +A^{\rm TAD}(a/r_0)
                                    +B^{\rm TAD}(a/r_0)^2.
                  \end{array}\right.
\label{eq_cont_ext}
\end{equation}
In other words, one requires that the data computed for
non-perturbative and tadpole improvement extrapolate to a common
continuum value. By combining the results obtained at three
$\beta$-values for tadpole improvement with those at two
$\beta$-values in the case of non-perturbative improvement we have
five data points to determine the four fit parameters $M, B^{\rm NP},
A^{\rm TAD}$ and~$B^{\rm TAD}$. Note that the data obtained on
$32^3\cdot64$ at $\beta=6.0$ have not been used. It is worth pointing
out that the spatial volume at $\beta=5.7$ is larger than those for
the larger two $\beta$-values which enter the extrapolations.

\begin{figure}
\vspace{1.0cm}
\setlength{\epsfxsize}{16cm}\epsfbox[18 350 554 769]{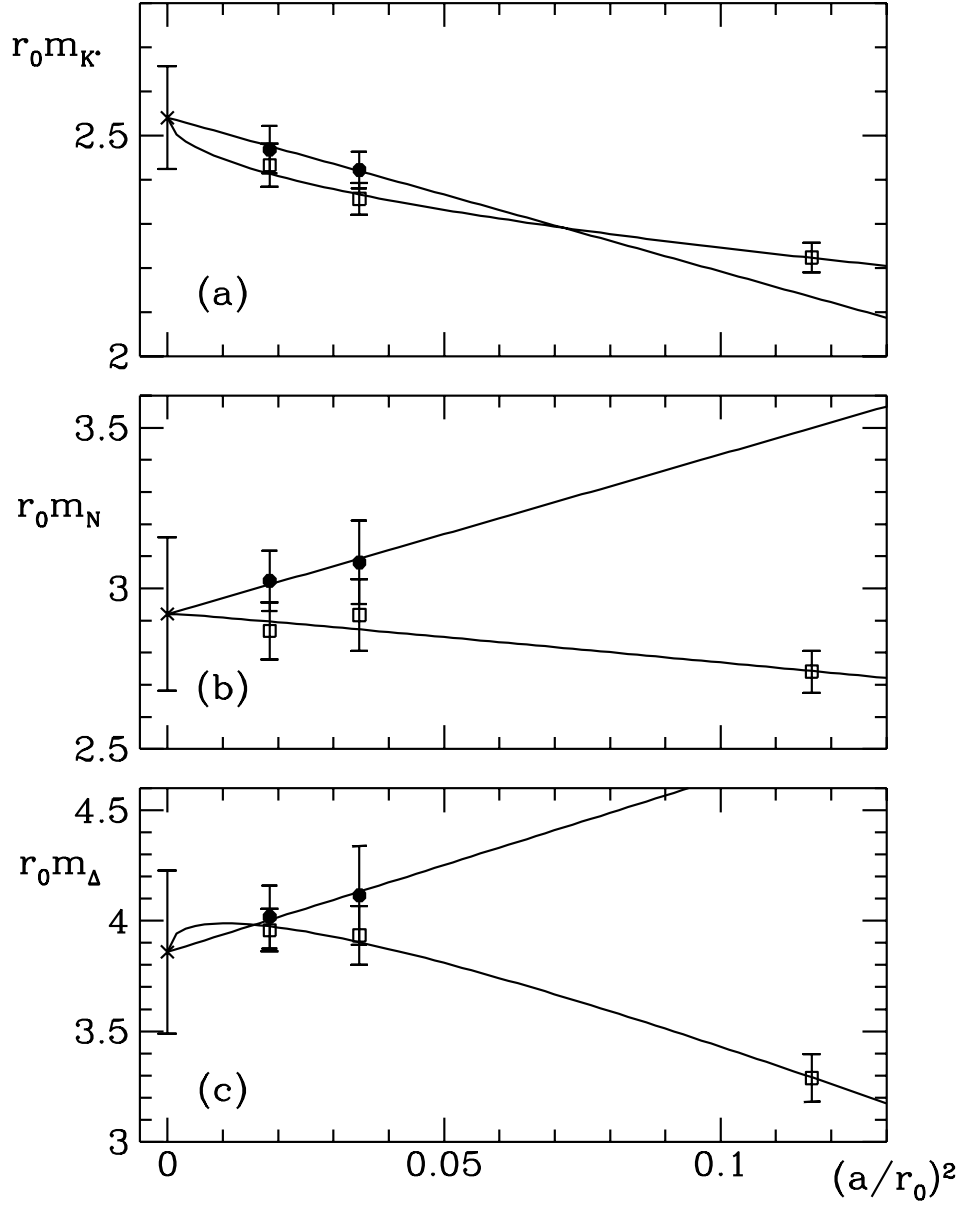}
\vspace{6.5cm}
\caption{Examples of simultaneous continuum extrapolations of the NP
and TAD datasets. (a): the $K^*$ meson, (b): the nucleon and (c): the
$\Delta$. Full circles denote the data computed using the
non-perturbative estimate of $\csw$, whereas the tadpole improved data
are represented by open squares.
\label{fig_cont_ext}}
\end{figure}

In Fig.~\ref{fig_cont_ext} we show examples of continuum
extrapolations based on eq.~(\ref{eq_cont_ext}), namely one
representative of each of the vector meson, octet and decuplet baryon
channels, respectively. The extrapolations have been repeated for the
other two choices of the lattice scale, i.e. $Q=m_{K^*}$ and~$m_N$.
The value of $\chi^2/\rm dof$ for these fits was quite low (below~1)
for all channels considered.

The results are listed in Tables~\ref{tab_res_r0}--\ref{tab_res_mnuc}
in the row labelled ``Cont.''. These numbers represent the final
results for the physical hadrons in this paper. In addition we have
also performed continuum extrapolations of hadrons interpolated to
$(\mps{r_0})^2=3.0$. The results are also included in
Table~\ref{tab_res_mref}. 

To check whether or not the continuum extrapolations for different
choices of the lattice scale are controlled, one can compare, for
instance, the continuum result for a particular hadron in units of
$m_{K^*}$, i.e. $(M/m_{K^*})$ with the ratio $(r_0M)/(r_0m_{K^*})$. It
then turns out that not only are the values consistent within errors,
but they are also numerically very close. This gives us further
confidence that the continuum estimates in
Tables~\ref{tab_res_r0}--\ref{tab_res_mnuc} are reliable.

From the results in the tables one can also estimate the size of
lattice artefacts at a fixed $\beta$-value, say, $\beta=6.0$ which
roughly corresponds to $a\approx0.1\,\fm$. Using the numbers from
Table~\ref{tab_res_r0} one infers that lattice artefacts for both
mesons and baryons at $\beta=6.0$ are of the order of 5\,\% or
less. Apart from Tables~\ref{tab_res_r0}--\ref{tab_res_mref} and
Fig.~\ref{fig_cont_ext} some information about the relative scaling
behaviour of the NP and TAD datasets can also be gained from
Figs.~\ref{fig_mpssq_vs_mqtilde}--\ref{fig_m_all_vs_mqtilde_tad}. Here
one observes that data for mesons and baryons in units of $r_0$ are
almost independent of the lattice spacing for $\beta\ge6.0$. This is
particularly pronounced when non-perturbative improvement is
employed. It is also clear that significant lattice artefacts are
present in the tadpole improved data at $\beta=5.7$.

\section{Discussion and conclusions}
\label{sec_concl}

Our final results in Tables~\ref{tab_res_r0}--\ref{tab_res_mnuc}
and~\ref{tab_res_mref} can now be compared to other lattice
calculations and experimental data.

In ref.~\cite{ALPHA_UKQCD} results for vector mesons computed using
the non-perturbative value of $\csw$ were presented. By comparing our
results for $r_0\mv$ computed at $(r_0\mps)^2=3.0$ and those for 
${r_0}m_{K^*}$ to ref.~\cite{ALPHA_UKQCD} we find differences of 1--2
standard deviations at most. Since the numerical procedures employed
in~\cite{ALPHA_UKQCD} are quite different to those used in this paper
(see also~\cite{ALPHA99}) this agreement is an important check of the
stability of our results, both at finite lattice spacing and in the
continuum limit.

\begin{figure}[tp]
\vspace{1.0cm}
\setlength{\epsfxsize}{16cm}\epsfbox[18 180 554 589]{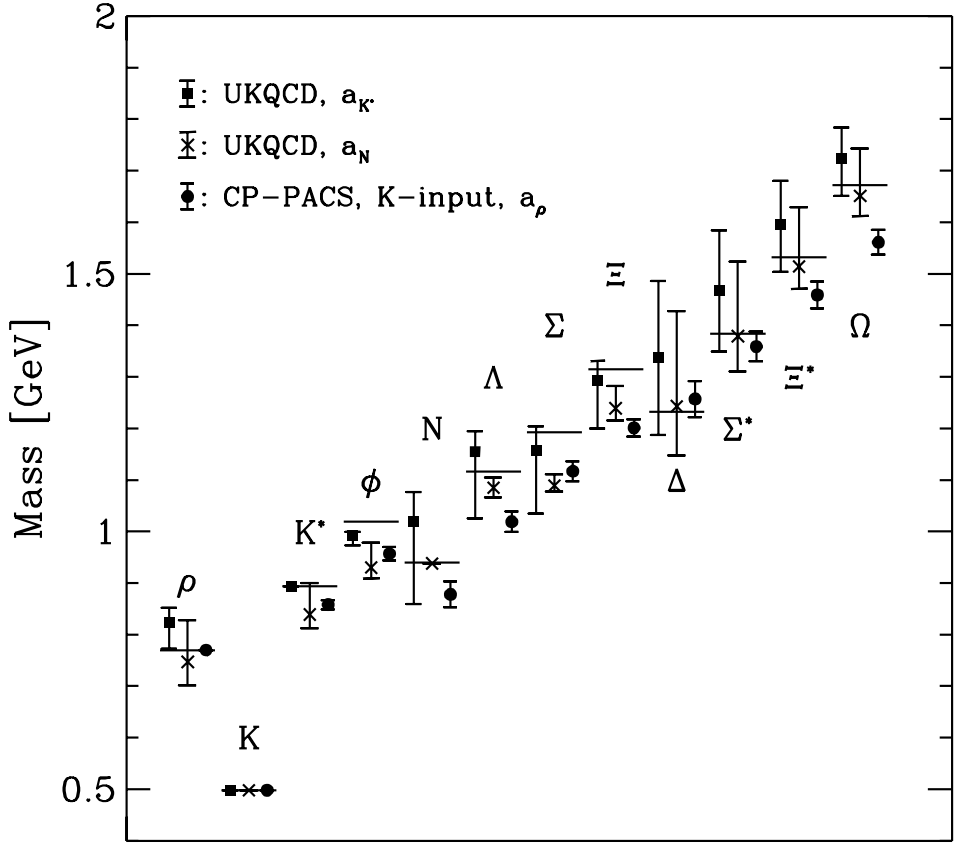}
%\vspace{-1.5cm}
\caption{The quenched light hadron spectrum computed in the $O(a)$
improved and its comparison to the results of
ref.~\protect\cite{CP-PACS_quen}, obtained using the unimproved Wilson
action (full circles). The levels of the experimental points are
denoted by the solid lines.
\label{fig_QuenchedSpectrum}}
\end{figure}

As mentioned in the introduction, one important goal of this study is
to corroborate earlier findings by GF11~\cite{GF11_quen} and the
CP-PACS Collaboration~\cite{CP-PACS_quen}, using an improved
discretisation of the QCD action. In Fig.~\ref{fig_QuenchedSpectrum}
we present our final results in physical units, computed using either
$m_{K^*}$ or $m_N$ to set the scale. Our numbers are compared to the
CP-PACS results, obtained using $m_\rho$ to set the scale, and to the
experimental numbers.

The first observation is that, on the whole, the two simulations agree
quite well, although the errors quoted by CP-PACS are in general much
smaller. This confirms the previous conclusion that the quenched light
hadron spectrum agrees within 10\,\% with experiment. This
independent confirmation, using results in the $O(a)$ improved theory,
is an important result, since lattice artefacts for unimproved Wilson
fermions are in general much larger~\cite{how_lat97,SCRI_imp}, so that
continuum extrapolations can be quite drastic.

It must be mentioned, though, that the overall precision of our
results cannot match that of ref.~\cite{CP-PACS_quen} for the
following reasons. Firstly, our calculations have been performed in a
fairly narrow range of quark masses, for relatively small volumes
(mostly for $L\approx1.5\,\fm$). Therefore, possible deviations from a
linear quark mass dependence close to the chiral limit could not be
detected or controlled. This affects mainly states like the $\rho$,
nucleon or the $\Delta$. Indeed, the modelling of the observed
downward curvature in the data reported in~\cite{CP-PACS_quen} turned
out to be a significant factor in the detection of the deviation from
the experimentally observed spectrum. Furthermore, as far as our
baryonic data are concerned, the chiral behaviour at $\beta=5.7$ is
not controlled at all, since only two data points have been computed,
so that we had to {\it assume} linearity in the quark mass. Finally,
our data are subject to finite-size corrections. For hadron masses
computed at the physical values of the quark masses these effects
amount to at most 2\,\% for vector mesons, $1-3\,\%$ for octet and
$4-8\,\%$ for decuplet baryons
(c.f. Tables~\ref{tab_res_r0}--\ref{tab_all_res_B60LARGE}).

To some extent one can check against possible finite-size effects in
our results by setting the scale using the nucleon mass. As a baryon
the nucleon might also be affected more by finite-size effects, and it
is reasonable to expect a (partial) cancellation of these effects in
the ratios $m_{\rm hadron}/m_N$. Indeed, as can be seen from
Fig.~\ref{fig_QuenchedSpectrum} our results in physical units decrease
when the scale is set by $m_N$ rather than $m_{K^*}$, although the
difference is mostly not significant.

If the differences in the results due to different choices of the
lattice scale is not attributed to finite-size effects, then they
serve to estimate the intrinsic scale ambiguity in the quenched
approximation in the continuum limit. On the basis of the results in
Tables~\ref{tab_res_r0}--\ref{tab_res_mnuc} one can infer that in the
most extreme cases the difference between the highest and lowest
values in physical units amounts to about 20\,\%. We may then assign
an uncertainty of $\pm10\,\%$ to our results in physical units as a
consequence of the scale ambiguity.

To summarise, we have presented a comprehensive study of the light
hadron spectrum in quenched QCD, using improved Wilson
actions. Qualitative results indicate an improved behaviour of the
pseudoscalar-vector mass splitting. The parameter~$J$, on the other
hand, shows no sign of approaching its phenomenological value
of~$0.48(2)$, even in the continuum limit. This appears to be an
intrinsic feature of the quenched approximation.

Our results show that $O(a)$ improvement works well for spectral
quantities. Hadron masses computed for non-perturbative~$\csw$ and
expressed in units of~$r_0$ show almost no dependence on the lattice
spacing for $\beta\ge6.0$. We also find that the extrapolations to the
continuum limit are quite mild in general. That is, for
$a\approx0.1\,\fm$ lattice artefacts amount to about 5\,\% or less.

We have presented further evidence that the quenched light hadron
spectrum agrees with experiment to within 10\,\%. Further technical
improvements, including, in particular, the modelling of the quark
mass dependence and the addition of more points in the continuum
extrapolations should be implemented to increase the overall
precision.

The next step is the extension of this investigation to decay
constants and matrix elements. First results have already been
presented in~\cite{rdk_lat96,par_lat9697,dlin_lpl_lat98}. Here, an
attractive feature is the availability of non-perturbative
determinations of some renormalisation
factors~\cite{ALPHA_ZA,ALPHA_quark}.

%%%%%%%%%%%%%%%%%%%%%%%%%%%%%%%%%%%%%%%%%%%%%%%%%%%%%%%%%%%%%%%%%%%%%%%
% \clearpage
\vskip 2mm
\noindent {\bf Acknowledgments.} This work was supported by the
Particle Physics \& Astronomy Research Council (PPARC) through grants
GR/L56336 and GR/L29927. We also acknowledge the support by EPSRC
under grant GR/K41663. H.P.S. acknowledges the support of the
Leverhulme foundation and the JSPS Research for the Future
program. HW is grateful to Akira Ukawa for the hospitality at the
Center of Computational Physics, University of Tsukuba, where part of
this work was completed. DGR and HW acknowledge the support by PPARC
through the award of Advanced Fellowships.
%%%%%%%%%%%%%%%%%%%%%%%%%%%%%%%%%%%%%%%%%%%%%%%%%%%%%%%%%%%%%%%%%%%%%%%

%%%%%%%%%%%%%%%%%%%%%%%%%%%%%%%%%%%%%%%%%%%%%%%%%%%%%%%%%%%%%%%%%%%%%%%

\begin{table}
\caption{Simulation parameters, statistics and smearing parameters for
the NP (upper three rows) and TAD (lower three rows) datasets. Lattice
sizes in physical units are estimated using $r_0$ to set the
scale~\protect\cite{alpha_r0}. The number of exceptional
configurations removed from the ensemble is denoted in parentheses.
\label{tab_simpar}
}
\begin{tabular}{ccccccl}
$\beta$ & $\csw$ & $L^3\cdot T$ & $L\,[\fm]$ & $\kappa$
        & {\#conf.} & smearing \\
\tableline
6.0 & 1.769 & $16^3\cdot48$ & 1.5 & $0.13344,\,0.13417,\,0.13455$ & 496(3)
    & fuzz, $r=6$ \\
    &       & $32^3\cdot64$ & 3.0 & $0.13344,\,0.13417,\,0.13455$ &  70(2)
    & jac, $N_{\rm jac}=30$  \\
6.2 & 1.614 & $24^3\cdot48$ & 1.6 & $0.13460,\,0.13510,\,0.13530$ & 216   
    & fuzz, $r=8$ \\
\tableline
5.7 & 1.568 & $16^3\cdot48$ & 2.7 & $0.13843,\,0.14077$           & 145
    & jac, $N_{\rm jac}=16$  \\
6.0 & 1.479 & $16^3\cdot48$ & 1.5 & $0.13700,\,0.13810,\,0.13856$ & 499
    & fuzz, $r=6$ \\
6.2 & 1.442 & $24^3\cdot48$ & 1.6 & $0.13640,\,0.13710,\,0.13745$ & 218
    & fuzz, $r=8$ \\
\end{tabular}
\end{table}

\begin{table}
\caption{Pseudoscalar masses for the non-perturbatively improved
data sets.
\label{tab_ps_NP}
}
\begin{tabular}{ccccr@{.}lcr@{.}l}
$\beta$ &$L^3\cdot T$ &$\kappa_{1}$ &$\kappa_{2}$ &
\multicolumn{2}{c}{$a\mps$} &$[t_{\rm min},\,t_{\rm max}]$
&\multicolumn{2}{c}{$\chi^2 /$ dof}\\  
\tableline
6.0 & $16^3\cdot 48$ & 0.13344 & 0.13344 & 0&3977$^{+13}_{-7}$ &
[ 6,23] & 23&82 / 30 \\  
 & & 0.13417 & 0.13344 & 0&3553$^{+15}_{-7}$  & [ 6,23] & 24&18 / 30 \\ 
 & & 0.13455 & 0.13344 & 0&3319$^{+17}_{-9}$  & [ 6,23] & 26&70 / 30 \\ 
 & & 0.13417 & 0.13417 & 0&3077$^{+18}_{-8}$  & [ 6,23] & 26&14 / 30 \\ 
 & & 0.13455 & 0.13417 & 0&2805$^{+19}_{-10}$ & [ 6,23] & 27&37 / 30 \\ 
 & & 0.13455 & 0.13455 & 0&2493$^{+22}_{-12}$ & [ 6,23] & 30&84 / 30 \\ 
 \tableline
6.0 & $32^3\cdot 64$ & 0.13344 & 0.13344 & 0&3952$^{+16}_{-8}$ &
[15,31] & 14&49 / 15 \\  
 & & 0.13417 & 0.13344 & 0&3524$^{+15}_{-10}$ & [15,31] & 14&86 / 15 \\ 
 & & 0.13455 & 0.13344 & 0&3284$^{+15}_{-11}$ & [15,31] & 13&56 / 15 \\ 
 & & 0.13417 & 0.13417 & 0&3048$^{+13}_{-11}$ & [15,31] & 13&64 / 15 \\ 
 & & 0.13455 & 0.13417 & 0&2769$^{+13}_{-11}$ & [15,31] & 12&22 / 15 \\ 
 & & 0.13455 & 0.13455 & 0&2457$^{+14}_{-10}$ & [15,31] & 13&04 / 15 \\ 
 \tableline
6.2 & $24^3\cdot 48$ & 0.13460 & 0.13460 & 0&2803$^{+15}_{-10}$ &
[ 8,23] & 30&99 / 26 \\  
 & & 0.13510 & 0.13460 & 0&2492$^{+17}_{-12}$ & [ 8,23] & 29&08 / 26 \\ 
 & & 0.13530 & 0.13460 & 0&2361$^{+18}_{-14}$ & [ 8,23] & 28&42 / 26 \\ 
 & & 0.13510 & 0.13510 & 0&2149$^{+19}_{-14}$ & [ 8,23] & 31&54 / 26 \\ 
 & & 0.13530 & 0.13510 & 0&1998$^{+19}_{-17}$ & [ 8,23] & 31&18 / 26 \\ 
 & & 0.13530 & 0.13530 & 0&1836$^{+23}_{-18}$ & [ 8,23] & 32&04 / 26 \\ 
\end{tabular}
\end{table}

\begin{table}
\caption{Vector masses for the non-perturbatively improved data sets.
\label{tab_vec_NP}}
\begin{tabular}{ccccr@{.}lcr@{.}l}
$\beta$ &$L^3\cdot T$ &$\kappa_{1}$ &$\kappa_{2}$ &
\multicolumn{2}{c}{$a\mv$} &$[t_{\rm min},\,t_{\rm max}]$
&\multicolumn{2}{c}{$\chi^2 /$ dof}\\  
\tableline
6.0 & $16^3\cdot 48$ & 0.13344 & 0.13344 & 0&5397$^{+32}_{-30}$ &
[ 6,23] & 24&02 / 30 \\  
 & & 0.13417 & 0.13344 & 0&5124$^{+49}_{-32}$ & [ 6,23] & 27&16 / 30 \\ 
 & & 0.13455 & 0.13344 & 0&4997$^{+51}_{-50}$ & [ 6,23] & 29&71 / 30 \\ 
 & & 0.13417 & 0.13417 & 0&4852$^{+53}_{-53}$ & [ 6,23] & 27&92 / 30 \\ 
 & & 0.13455 & 0.13417 & 0&4713$^{+69}_{-68}$ & [ 6,23] & 31&96 / 30 \\ 
 & & 0.13455 & 0.13455 & 0&4577$^{+85}_{-83}$ & [ 6,23] & 30&16 / 30 \\ 
 \tableline
6.0 & $32^3\cdot 64$ & 0.13344 & 0.13344 & 0&5400$^{+47}_{-35}$ &
[10,20] & 13&68 / 9 \\ 
 & & 0.13417 & 0.13344 & 0&5143$^{+53}_{-39}$ & [10,20] & 14&44 / 9 \\ 
 & & 0.13455 & 0.13344 & 0&5019$^{+58}_{-46}$ & [10,20] & 13&75 / 9 \\ 
 & & 0.13417 & 0.13417 & 0&4887$^{+61}_{-48}$ & [10,20] & 13&30 / 9 \\ 
 & & 0.13455 & 0.13417 & 0&4762$^{+74}_{-59}$ & [10,20] & 10&31 / 9 \\ 
 & & 0.13455 & 0.13455 & 0&4636$^{+88}_{-76}$ & [10,20] &  6&69 / 9 \\ 
 \tableline
6.2 & $24^3\cdot 48$ & 0.13460 & 0.13460 & 0&3887$^{+32}_{-28}$ &
[ 8,23] & 33&55 / 26 \\  
 & & 0.13510 & 0.13460 & 0&3708$^{+42}_{-36}$ & [ 8,23] & 28&99 / 26 \\ 
 & & 0.13530 & 0.13460 & 0&3645$^{+43}_{-47}$ & [ 8,23] & 26&51 / 26 \\ 
 & & 0.13510 & 0.13510 & 0&3531$^{+55}_{-51}$ & [ 8,23] & 29&56 / 26 \\ 
 & & 0.13530 & 0.13510 & 0&3471$^{+62}_{-61}$ & [ 8,23] & 27&91 / 26 \\ 
 & & 0.13530 & 0.13530 & 0&3414$^{+72}_{-82}$ & [ 8,23] & 30&98 / 26 \\ 
\end{tabular}
\end{table}

\begin{table}
\caption{Pseudoscalar masses for the tadpole improved data sets.
\label{tab_ps_TAD}}
\begin{tabular}{ccccr@{.}lcc}
$\beta$ &$L^3\cdot T$ &$\kappa_{1}$ &$\kappa_{2}$
&\multicolumn{2}{c}{$a\mps$} &$[t_{\rm min},\,t_{\rm max}]$ &$\chi^2 /$ dof\\ 
\tableline 
5.7 & $16^3\cdot 32$ & 0.13843 & 0.13843 & 0&7350$^{+11}_{-6}$ &
[ 6,15] & 23.99 / 14 \\  
 & & 0.14077 & 0.13843 & 0&6404$^{+11}_{-29}$ & [ 5,15] & 16.89 / 16 \\ 
 & & 0.14077 & 0.14077 & 0&5307$^{+19}_{-20}$ & [ 5,15] & 16.00 / 16 \\ 
 \tableline
6.0 & $16^3\cdot 48$ & 0.13700 & 0.13700 & 0&4131$^{+13}_{-7}$
& [ 6,23] & 22.57 / 30 \\  
 & & 0.13810 & 0.13700 & 0&3572$^{+16}_{-6}$ & [ 6,23] & 21.97 / 30 \\ 
 & & 0.13856 & 0.13700 & 0&3320$^{+21}_{-6}$ & [ 6,23] & 27.82 / 30 \\ 
 & & 0.13810 & 0.13810 & 0&2927$^{+21}_{-4}$ & [ 6,23] & 26.93 / 30 \\ 
 & & 0.13856 & 0.13810 & 0&2621$^{+23}_{-7}$ & [ 6,23] & 29.93 / 30 \\ 
 & & 0.13856 & 0.13856 & 0&2268$^{+25}_{-10}$ & [ 6,23] & 30.40 / 30 \\ 
 \tableline
6.2 & $24^3\cdot 48$ & 0.13640 & 0.13640 & 0&3033$^{+12}_{-10}$
& [ 8,23] & 30.03 / 26 \\  
 & & 0.13710 & 0.13640 & 0&2643$^{+15}_{-11}$ & [ 8,23] & 29.29 / 26 \\ 
 & & 0.13745 & 0.13640 & 0&2436$^{+18}_{-13}$ & [ 8,23] & 29.02 / 26 \\ 
 & & 0.13710 & 0.13710 & 0&2206$^{+18}_{-12}$ & [ 8,23] & 31.97 / 26 \\ 
 & & 0.13745 & 0.13710 & 0&1959$^{+21}_{-16}$ & [ 8,23] & 30.92 / 26 \\ 
 & & 0.13745 & 0.13745 & 0&1680$^{+27}_{-18}$ & [ 8,23] & 31.13 / 26 \\ 
\end{tabular}
\end{table}

\begin{table}
\caption{Vector masses for the tadpole improved data sets.
\label{tab_vec_TAD}}
\begin{tabular}{ccccr@{.}lcc}
$\beta$ &$L^3\cdot T$ &$\kappa_{1}$ &$\kappa_{2}$
&\multicolumn{2}{c}{$a\mv$} &$[t_{\rm min},\,t_{\rm max}]$ &$\chi^2 /$ dof\\ 
\tableline
5.7 & $16^3\cdot 32$ & 0.13843 & 0.13843 & 0&9332$^{+45}_{-37}$ &
[ 7,15] & 12.80 / 12 \\  
 & & 0.14077 & 0.13843 & 0&8688$^{+54}_{-48}$ & [ 7,15] & 11.96 / 12 \\ 
 & & 0.14077 & 0.14077 & 0&809$^{+9}_{-13}$ & [ 7,15] & 10.76 / 12 \\ 
 \tableline
6.0 & $16^3\cdot 48$ & 0.13700 & 0.13700 & 0&5386$^{+32}_{-22}$ &
[ 7,23] & 23.82 / 28 \\  
 & & 0.13810 & 0.13700 & 0&5030$^{+40}_{-24}$ & [ 6,23] & 27.20 / 30 \\ 
 & & 0.13856 & 0.13700 & 0&4889$^{+45}_{-41}$ & [ 6,23] & 27.83 / 30 \\ 
 & & 0.13810 & 0.13810 & 0&4652$^{+52}_{-44}$ & [ 6,23] & 26.92 / 30 \\ 
 & & 0.13856 & 0.13810 & 0&4501$^{+66}_{-62}$ & [ 6,23] & 29.23 / 30 \\ 
 & & 0.13856 & 0.13856 & 0&4353$^{+88}_{-76}$ & [ 6,23] & 25.43 / 30 \\ 
 \tableline
6.2 & $24^3\cdot 48$ & 0.13640 & 0.13640 & 0&4005$^{+23}_{-26}$ &
[ 8,23] & 32.65 / 26 \\  
 & & 0.13710 & 0.13640 & 0&3761$^{+34}_{-29}$ & [ 8,23] & 28.79 / 26 \\ 
 & & 0.13745 & 0.13640 & 0&3648$^{+39}_{-44}$ & [ 8,23] & 25.39 / 26 \\ 
 & & 0.13710 & 0.13710 & 0&3522$^{+50}_{-44}$ & [ 8,23] & 26.83 / 26 \\ 
 & & 0.13745 & 0.13710 & 0&3412$^{+64}_{-63}$ & [ 8,23] & 24.85 / 26 \\ 
 & & 0.13745 & 0.13745 & 0&3306$^{+90}_{-95}$ & [ 8,23] & 28.83 / 26 \\ 
\end{tabular}
\end{table}

\begin{table}
\caption{Masses for the nucleon and $\Delta$ for degenerate quark
mass combinations for the non-perturbatively improved data.
\label{tab_bardeg_NP}}
\begin{tabular}{cccr@{.}lccr@{.}lcc}
$\beta$ &$L^3\cdot T$ &$\kappa$
&\multicolumn{2}{c}{$am_{\rm N}$} &$[t_{\rm min},\,t_{\rm max}]$
&$\chi^2 /$ dof 
&\multicolumn{2}{c}{$am_\Delta$}  &$[t_{\rm min},\,t_{\rm max}]$
&$\chi^2 /$ dof\\  
\tableline
6.0 & $16^3\cdot 48$ 
 & 0.13344 & 0&808$^{+10}_{-7}$  & [ 9,23] & 25.42/24
           & 0&913$^{+9}_{-22}$  & [ 9,23] & 21.84/24 \\
&& 0.13417 & 0&711$^{+16}_{-13}$ & [ 9,23] & 25.74/24
           & 0&852$^{+19}_{-23}$ & [ 9,23] & 31.33/24 \\
&& 0.13455 & 0&665$^{+26}_{-28}$ & [ 9,23] & 26.98/24
           & 0&768$^{+52}_{-36}$ & [10,23] & 40.37/22 \\
 \tableline
6.0 & $32^3\cdot 64$ 
 & 0.13344 & 0&799$^{+10}_{-10}$ & [ 3,18] & 16.82/12
           & 0&899$^{+13}_{-14}$ & [ 2,16] & 17.56/11 \\
&& 0.13417 & 0&700$^{+11}_{-15}$ & [ 3,18] & 15.99/12
           & 0&818$^{+16}_{-13}$ & [ 2,16] & 20.09/11 \\
&& 0.13455 & 0&641$^{+16}_{-20}$ & [ 3,18] & 14.63/12
           & 0&781$^{+18}_{-14}$ & [ 2,16] & 21.66/11 \\
 \tableline
6.2 & $24^3\cdot 48$
 & 0.13460 & 0&586$^{+8}_{-6}$   & [10,23] & 42.46/22
           & 0&671$^{+8}_{-7}$   & [11,23] & 20.35/20 \\
&& 0.13510 & 0&509$^{+10}_{-10}$ & [10,23] & 41.09/22
           & 0&618$^{+14}_{-12}$ & [11,23] & 21.54/20 \\
&& 0.13530 & 0&487$^{+3}_{-14}$  & [10,23] & 30.35/22
           & 0&596$^{+19}_{-13}$ & [11,23] & 20.31/20 \\
\end{tabular}
\end{table}

\begin{table}
\caption{Masses for the nucleon and $\Delta$ for degenerate quark
mass combinations for the tadpole improved data.
\label{tab_bardeg_TAD}}
\begin{tabular}{cccr@{.}lccr@{.}lcc}
$\beta$ &$L^3\cdot T$ &$\kappa$
&\multicolumn{2}{c}{$am_{\rm N}$} &$[t_{\rm min},\,t_{\rm max}]$
&$\chi^2 /$ dof 
&\multicolumn{2}{c}{$am_\Delta$}  &$[t_{\rm min},\,t_{\rm max}]$
&$\chi^2 /$ dof\\  
\tableline
5.7 & $16^3\cdot 32$ 
 & 0.13843 & 1&423$^{+12}_{-4}$  &  [ 7,15] &  7.37/12
           & 1&539$^{+21}_{-8}$  &  [ 7,15] & 23.62/12 \\
&& 0.14077 & 1&183$^{+14}_{-11}$ &  [ 6,15] & 15.96/14
           & 1&334$^{+26}_{-17}$ &  [ 7,15] & 10.59/12 \\
\tableline
6.0 & $16^3\cdot 48$ 
 & 0.13700 & 0&817$^{+9}_{-4}$   &  [10,23] & 21.26/22
           & 0&909$^{+7}_{-9}$   &  [10,23] & 23.49/22 \\
&& 0.13810 & 0&678$^{+17}_{-8}$  &  [10,23] & 27.46/22
           & 0&810$^{+13}_{-13}$ &  [ 8,23] & 23.95/26 \\
&& 0.13856 & 0&616$^{+26}_{-16}$ &  [10,23] & 26.95/22
           & 0&774$^{+21}_{-26}$ &  [ 8,23] & 41.53/26 \\
 \tableline
6.2 & $24^3\cdot 48$
 & 0.13640 & 0&608$^{+8}_{-6}$   &  [11,23] & 34.87/20
           & 0&691$^{+7}_{-7}$   &  [11,23] & 19.42/20 \\
&& 0.13710 & 0&509$^{+12}_{-9}$  &  [11,23] & 38.62/20
           & 0&620$^{+11}_{-10}$ &  [11,23] & 23.87/20 \\
&& 0.13745 & 0&467$^{+12}_{-20}$ &  [11,23] & 22.68/20
           & 0&577$^{+19}_{-13}$ &  [11,23] & 20.60/20 \\
\end{tabular}
\end{table}

\begin{table}
\caption{Masses for $\Sigma$-like, $\Lambda$-like and $\Delta$-like
baryons for non-degenerate quark masses for $\beta=6.0$, $L^3\cdot
T=16^3\cdot 48$ for the non-perturbatively improved data.
\label{tab_barnondeg_NP_60small}}
\begin{tabular}{cccr@{.}lr@{.}lr@{.}l}
$\kappa_1$ & $\kappa_2$ & $\kappa_3$ 
&\multicolumn{2}{c}{$am_\Sigma$} 
&\multicolumn{2}{c}{$am_\Lambda$}
&\multicolumn{2}{c}{$am_\Delta$}  \\
\tableline
 0.13344 & 0.13344 & 0.13417
 & 0&780$^{+10}_{-6}$  & 0&775$^{+11}_{-6}$  & 0&894$^{+11}_{-12}$ \\
 0.13344 & 0.13344 & 0.13455
 & 0&766$^{+11}_{-7}$  & 0&757$^{+12}_{-7}$  & 0&890$^{+14}_{-12}$ \\
 0.13344 & 0.13417 & 0.13417
 & 0&743$^{+12}_{-8}$  & 0&744$^{+13}_{-7}$  & 0&872$^{+14}_{-13}$ \\
 0.13344 & 0.13417 & 0.13455
 & 0&735$^{+12}_{-10}$ & 0&722$^{+15}_{-10}$ & 0&871$^{+15}_{-14}$ \\
 0.13344 & 0.13455 & 0.13455
 & 0&708$^{+13}_{-14}$ & 0&713$^{+16}_{-11}$ & 0&860$^{+19}_{-18}$ \\
 0.13417 & 0.13344 & 0.13344
 & 0&771$^{+11}_{-6}$  & 0&781$^{+11}_{-6}$  & \multicolumn{2}{c}{$\mbox{}$}\\
 0.13417 & 0.13344 & 0.13417
 & 0&745$^{+14}_{-7}$  & 0&739$^{+12}_{-8}$  & \multicolumn{2}{c}{$\mbox{}$}\\
 0.13417 & 0.13344 & 0.13455
 & 0&728$^{+12}_{-10}$ & 0&727$^{+14}_{-9}$  & \multicolumn{2}{c}{$\mbox{}$}\\
 0.13417 & 0.13417 & 0.13455
 & 0&697$^{+13}_{-13}$ & 0&692$^{+16}_{-12}$ & 0&845$^{+21}_{-20}$ \\
 0.13417 & 0.13455 & 0.13455
 & 0&677$^{+15}_{-17}$ & 0&679$^{+17}_{-15}$ & 0&837$^{+25}_{-27}$ \\
 0.13455 & 0.13344 & 0.13344
 & 0&750$^{+13}_{-7}$  & 0&769$^{+12}_{-8}$  & \multicolumn{2}{c}{$\mbox{}$}\\
 0.13455 & 0.13344 & 0.13417
 & 0&719$^{+15}_{-10}$ & 0&734$^{+13}_{-10}$ & \multicolumn{2}{c}{$\mbox{}$}\\
 0.13455 & 0.13344 & 0.13455
 & 0&717$^{+15}_{-12}$ & 0&702$^{+14}_{-14}$ & \multicolumn{2}{c}{$\mbox{}$}\\
 0.13455 & 0.13417 & 0.13417
 & 0&689$^{+18}_{-13}$ & 0&698$^{+14}_{-13}$ & \multicolumn{2}{c}{$\mbox{}$}\\
 0.13455 & 0.13417 & 0.13455
 & 0&679$^{+18}_{-15}$ & 0&674$^{+15}_{-19}$ & \multicolumn{2}{c}{$\mbox{}$}\\
\end{tabular}
\end{table}

\begin{table}
\caption{Masses for $\Sigma$-like, $\Lambda$-like and $\Delta$-like
baryons for non-degenerate quark masses for $\beta=6.0$, $L^3\cdot
T=32^3\cdot 64$ for the non-perturbatively improved data.
\label{tab_barnondeg_NP_60large}}
\begin{tabular}{cccr@{.}lr@{.}lr@{.}l}
$\kappa_1$ & $\kappa_2$ & $\kappa_3$ 
&\multicolumn{2}{c}{$am_\Sigma$} 
&\multicolumn{2}{c}{$am_\Lambda$}
&\multicolumn{2}{c}{$am_\Delta$}  \\
 \tableline
 0.13344 & 0.13344 & 0.13417
 & 0&769$^{+10}_{-11}$ & 0&765$^{+10}_{-11}$ &  0&873$^{+14}_{-14}$  \\
 0.13344 & 0.13344 & 0.13455
 & 0&755$^{+10}_{-12}$ & 0&748$^{+10}_{-13}$ &  0&859$^{+15}_{-13}$ \\
 0.13344 & 0.13417 & 0.13417
 & 0&732$^{+10}_{-13}$ & 0&737$^{+10}_{-12}$ &  0&845$^{+15}_{-14}$ \\
 0.13344 & 0.13417 & 0.13455
 & 0&726$^{+10}_{-13}$ & 0&711$^{+10}_{-15}$ &  0&832$^{+16}_{-13}$ \\
 0.13344 & 0.13455 & 0.13455
 & 0&695$^{+12}_{-16}$ & 0&704$^{+10}_{-15}$ &  0&820$^{+17}_{-13}$ \\
 0.13417 & 0.13344 & 0.13344
 & 0&764$^{+9}_{-11}$  & 0&772$^{+10}_{-11}$ & \multicolumn{2}{c}{$\mbox{}$}\\
 0.13417 & 0.13344 & 0.13417
 & 0&740$^{+10}_{-12}$ & 0&730$^{+10}_{-13}$ & \multicolumn{2}{c}{$\mbox{}$}\\
 0.13417 & 0.13344 & 0.13455
 & 0&716$^{+10}_{-14}$ & 0&719$^{+11}_{-14}$ & \multicolumn{2}{c}{$\mbox{}$}\\
 0.13417 & 0.13417 & 0.13455
 & 0&684$^{+12}_{-16}$ & 0&680$^{+12}_{-16}$ &  0&805$^{+17}_{-13}$ \\
 0.13417 & 0.13455 & 0.13455
 & 0&659$^{+15}_{-19}$ & 0&664$^{+13}_{-18}$ &  0&793$^{+18}_{-14}$ \\
 0.13455 & 0.13344 & 0.13344
 & 0&745$^{+10}_{-13}$ & 0&759$^{+10}_{-11}$ & \multicolumn{2}{c}{$\mbox{}$}\\
 0.13455 & 0.13344 & 0.13417
 & 0&712$^{+11}_{-14}$ & 0&724$^{+10}_{-14}$ & \multicolumn{2}{c}{$\mbox{}$}\\
 0.13455 & 0.13344 & 0.13455
 & 0&711$^{+10}_{-15}$ & 0&691$^{+12}_{-16}$ & \multicolumn{2}{c}{$\mbox{}$}\\
 0.13455 & 0.13417 & 0.13417
 & 0&678$^{+13}_{-16}$ & 0&684$^{+10}_{-17}$ & \multicolumn{2}{c}{$\mbox{}$}\\
 0.13455 & 0.13417 & 0.13455
 & 0&666$^{+13}_{-18}$ & 0&657$^{+16}_{-19}$ & \multicolumn{2}{c}{$\mbox{}$}\\
\end{tabular}
\end{table}

\begin{table}
\caption{Masses for $\Sigma$-like, $\Lambda$-like and $\Delta$-like
baryons for non-degenerate quark masses for $\beta=6.2$, $L^3\cdot
T=24^3\cdot 48$ for the non-perturbatively improved data.
\label{tab_barnondeg_NP_62}}
\begin{tabular}{cccr@{.}lr@{.}lr@{.}l}
$\kappa_1$ & $\kappa_2$ & $\kappa_3$
&\multicolumn{2}{c}{$am_\Sigma$} 
&\multicolumn{2}{c}{$am_\Lambda$}
&\multicolumn{2}{c}{$am_\Delta$}  \\
 \tableline
 0.13460 & 0.13460 & 0.13510
 & 0&555$^{+3}_{-9}$  & 0&548$^{+8}_{-7}$   & 0&656$^{+7}_{-8}$   \\
 0.13460 & 0.13460 & 0.13530
 & 0&547$^{+1}_{-10}$ & 0&534$^{+9}_{-8}$   & 0&648$^{+8}_{-8}$   \\
 0.13460 & 0.13510 & 0.13510
 & 0&528$^{+8}_{-10}$ & 0&524$^{+9}_{-8}$   & 0&638$^{+9}_{-8}$   \\
 0.13460 & 0.13510 & 0.13530
 & 0&522$^{+10}_{-10}$& 0&510$^{+7}_{-11}$  & 0&630$^{+10}_{-8}$  \\
 0.13460 & 0.13530 & 0.13530
 & 0&513$^{+7}_{-12}$ & 0&502$^{+8}_{-12}$  & 0&623$^{+11}_{-9}$  \\
 0.13510 & 0.13460 & 0.13460
 & 0&545$^{+9}_{-6}$  & 0&552$^{+9}_{-7}$   & \multicolumn{2}{c}{$\mbox{}$}\\
 0.13510 & 0.13460 & 0.13510
 & 0&525$^{+9}_{-8}$  & 0&522$^{+9}_{-10}$  & \multicolumn{2}{c}{$\mbox{}$}\\
 0.13510 & 0.13460 & 0.13530
 & 0&514$^{+9}_{-10}$ & 0&516$^{+7}_{-10}$  & \multicolumn{2}{c}{$\mbox{}$}\\
 0.13510 & 0.13510 & 0.13530
 & 0&500$^{+6}_{-11}$ & 0&497$^{+3}_{-11}$  & 0&611$^{+12}_{-10}$ \\
 0.13510 & 0.13530 & 0.13530
 & 0&497$^{+7}_{-11}$ & 0&474$^{+4}_{-27}$  & 0&606$^{+13}_{-11}$ \\
 0.13530 & 0.13460 & 0.13460
 & 0&530$^{+9}_{-7}$  & 0&543$^{+10}_{-7}$  & \multicolumn{2}{c}{$\mbox{}$}\\
 0.13530 & 0.13460 & 0.13510
 & 0&509$^{+7}_{-11}$ & 0&516$^{+11}_{-10}$ & \multicolumn{2}{c}{$\mbox{}$}\\
 0.13530 & 0.13460 & 0.13530
 & 0&503$^{+9}_{-11}$ & 0&507$^{+6}_{-14}$  & \multicolumn{2}{c}{$\mbox{}$}\\
 0.13530 & 0.13510 & 0.13510
 & 0&493$^{+1}_{-12}$ & 0&497$^{+7}_{-13}$  & \multicolumn{2}{c}{$\mbox{}$}\\
 0.13530 & 0.13510 & 0.13530
 & 0&487$^{+1}_{-13}$ & 0&497$^{+4}_{-15}$  & \multicolumn{2}{c}{$\mbox{}$}\\
\end{tabular}
\end{table}

\begin{table}
\caption{Masses for $\Sigma$-like, $\Lambda$-like and $\Delta$-like
baryons for non-degenerate quark masses for $\beta=6.0$, $L^3\cdot
T=16^3\cdot 48$ for the tadpole improved data.
\label{tab_barnondeg_TAD_60}}
\begin{tabular}{cccr@{.}lr@{.}lr@{.}l}
$\kappa_1$ & $\kappa_2$ & $\kappa_3$ 
&\multicolumn{2}{c}{$am_\Sigma$} 
&\multicolumn{2}{c}{$am_\Lambda$}
&\multicolumn{2}{c}{$am_\Delta$}  \\
\tableline
 0.13700 & 0.13700 & 0.13810
  & 0&777$^{+11}_{-5}$ & 0&771$^{+11}_{-5}$  & 0&873$^{+8}_{-9}$  \\
 0.13700 & 0.13700 & 0.13856
  & 0&761$^{+12}_{-5}$ & 0&749$^{+13}_{-6}$  & 0&882$^{+6}_{-15}$ \\
 0.13700 & 0.13810 & 0.13810
  & 0&722$^{+14}_{-7}$ & 0&732$^{+13}_{-7}$  & 0&853$^{+4}_{-18}$ \\
 0.13700 & 0.13810 & 0.13856
  & 0&713$^{+15}_{-8}$ & 0&703$^{+16}_{-9}$  & 0&845$^{+9}_{-17}$ \\
 0.13700 & 0.13856 & 0.13856
  & 0&678$^{+19}_{-12}$ & 0&697$^{+15}_{-9}$ & 0&832$^{+12}_{-17}$\\
 0.13810 & 0.13700 & 0.13700
  & 0&766$^{+11}_{-5}$ & 0&778$^{+12}_{-5}$  & \multicolumn{2}{c}{$\mbox{}$}\\
 0.13810 & 0.13700 & 0.13810
  & 0&733$^{+14}_{-7}$ & 0&718$^{+14}_{-7}$ & \multicolumn{2}{c}{$\mbox{}$}\\
 0.13810 & 0.13700 & 0.13856
  & 0&706$^{+14}_{-7}$ & 0&707$^{+15}_{-8}$ & \multicolumn{2}{c}{$\mbox{}$}\\
 0.13810 & 0.13810 & 0.13856
 & 0&659$^{+18}_{-13}$ & 0&659$^{+20}_{-12}$ & 0&810$^{+12}_{-20}$\\ 
 0.13810 & 0.13856 & 0.13856
 & 0&634$^{+22}_{-19}$ & 0&643$^{+20}_{-12}$ & 0&787$^{+19}_{-19}$\\ 
 0.13856 & 0.13700 & 0.13700
  & 0&741$^{+13}_{-7}$ & 0&763$^{+14}_{-5}$  & \multicolumn{2}{c}{$\mbox{}$}\\
 0.13856 & 0.13700 & 0.13810
  & 0&698$^{+17}_{-8}$ & 0&710$^{+16}_{-7}$  & \multicolumn{2}{c}{$\mbox{}$}\\
 0.13856 & 0.13700 & 0.13856
 & 0&697$^{+15}_{-8}$ & 0&672$^{+18}_{-12}$  & \multicolumn{2}{c}{$\mbox{}$}\\
 0.13856 & 0.13810 & 0.13810
 & 0&654$^{+21}_{-11}$ & 0&655$^{+18}_{-12}$ & \multicolumn{2}{c}{$\mbox{}$}\\
 0.13856 & 0.13810 & 0.13856
 & 0&642$^{+21}_{-10}$ & 0&626$^{+23}_{-19}$ & \multicolumn{2}{c}{$\mbox{}$}\\
\end{tabular}
\end{table}

\begin{table}
\caption{Masses for $\Sigma$-like, $\Lambda$-like and $\Delta$-like
baryons for non-degenerate quark masses for $\beta=6.2$, $L^3\cdot
T=24^3\cdot 48$ for the tadpole improved data.
\label{tab_barnondeg_TAD_62}}
\begin{tabular}{cccr@{.}lr@{.}lr@{.}l}
$\kappa_1$ & $\kappa_2$ & $\kappa_3$ 
&\multicolumn{2}{c}{$am_\Sigma$} 
&\multicolumn{2}{c}{$am_\Lambda$}
&\multicolumn{2}{c}{$am_\Delta$}  \\
\tableline
 0.13640 & 0.13640 & 0.13710
   & 0&576$^{+3}_{-8}$ & 0&568$^{+7}_{-6}$  & 0&656$^{+7}_{-6}$  \\
 0.13640 & 0.13640 & 0.13745
   & 0&554$^{+6}_{-7}$ & 0&545$^{+9}_{-7}$ & 0&652$^{+8}_{-7}$  \\
 0.13640 & 0.13710 & 0.13710
  & 0&539$^{+9}_{-8}$  & 0&535$^{+11}_{-7}$ & 0&642$^{+9}_{-7}$  \\
 0.13640 & 0.13710 & 0.13745
   & 0&526$^{+9}_{-8}$ & 0&510$^{+9}_{-9}$  & 0&627$^{+10}_{-8}$ \\
 0.13640 & 0.13745 & 0.13745
 & 0&504$^{+9}_{-13}$ & 0&491$^{+16}_{-10}$  & 0&619$^{+11}_{-9}$ \\
 0.13710 & 0.13640 & 0.13640
   & 0&563$^{+6}_{-5}$ & 0&571$^{+8}_{-7}$  & \multicolumn{2}{c}{$\mbox{}$}\\
 0.13710 & 0.13640 & 0.13710
  & 0&539$^{+7}_{-8}$ & 0&532$^{+11}_{-8}$  & \multicolumn{2}{c}{$\mbox{}$}\\
 0.13710 & 0.13640 & 0.13745
   & 0&515$^{+10}_{-8}$ & 0&518$^{+8}_{-8}$ & \multicolumn{2}{c}{$\mbox{}$}\\
 0.13710 & 0.13710 & 0.13745
  & 0&494$^{+6}_{-11}$ & 0&491$^{+3}_{-11}$  & 0&602$^{+11}_{-10}$\\
 0.13710 & 0.13745 & 0.13745
  & 0&463$^{+12}_{-24}$ & 0&457$^{+5}_{-22}$ & 0&593$^{+12}_{-11}$\\
 0.13745 & 0.13640 & 0.13640
   & 0&540$^{+6}_{-7}$ & 0&558$^{+9}_{-7}$  & \multicolumn{2}{c}{$\mbox{}$}\\
 0.13745 & 0.13640 & 0.13710
  & 0&510$^{+10}_{-8}$ & 0&522$^{+10}_{-9}$ & \multicolumn{2}{c}{$\mbox{}$}\\
 0.13745 & 0.13640 & 0.13745
  & 0&497$^{+15}_{-9}$ & 0&492$^{+8}_{-15}$ & \multicolumn{2}{c}{$\mbox{}$}\\
 0.13745 & 0.13710 & 0.13710
  & 0&470$^{+8}_{-19}$ & 0&492$^{+7}_{-13}$ & \multicolumn{2}{c}{$\mbox{}$}\\
 0.13745 & 0.13710 & 0.13745
  & 0&455$^{+7}_{-23}$ & 0&451$^{+13}_{-25}$ & \multicolumn{2}{c}{$\mbox{}$}\\
\end{tabular}
\end{table}

\begin{table}
\caption{Values for $\kcrit$ for all datasets. The labels `tree',
`bare' and `tad' refer to the coefficient $\bm$ estimated either at
tree-level, bare and tadpole improved one-loop perturbation theory,
respectively. Also shown are the results of
ref.~\protect\cite{AlphaIII}, obtained using the current quark mass
instead of the pseudoscalar mass.
\label{tab_kappac}}
\begin{tabular}{lccccccc}
 & $\beta$ & $L^3\cdot T$ & $\bm=0$ & tree & bare & tad 
 & ref.~\protect\cite{AlphaIII} \\
\tableline
NP  & 6.0  & $16^3\cdot48$ &
           & 0.135259$^{+16}_{-9}$  & 0.135255$^{+16}_{-9}$ 
           & 0.135252$^{+16}_{-9}$  &  \\
    & 6.0  & $32^3\cdot64$ &
           & 0.135241$^{+9}_{-10}$  & 0.135237$^{+9}_{-10}$
           & 0.135235$^{+9}_{-10}$  & \rb{0.135196(14)} \\
    & 6.2  & $24^3\cdot48$ &
           & 0.135818$^{+17}_{-14}$ & 0.135816$^{+17}_{-14}$
           & 0.135815$^{+17}_{-14}$ & 0.135795(13) \\
\tableline
TAD & 5.7  & $16^3\cdot48$ & 0.143408$^{+29}_{-45}$
           & 0.143240$^{+27}_{-40}$ & 0.143206$^{+26}_{-39}$
           & 0.143179$^{+26}_{-38}$ & \\
    & 6.0  & $16^3\cdot48$ & 0.139240$^{+20}_{-7}$
           & 0.139216$^{+19}_{-6}$  & 0.139212$^{+19}_{-6}$
           & 0.139209$^{+19}_{-6}$ & \\
    & 6.2  & $24^3\cdot48$ & 0.137912$^{+19}_{-13}$
           & 0.137900$^{+18}_{-12}$ & 0.137898$^{+18}_{-12}$
           & 0.137897$^{+18}_{-12}$ & \\
\end{tabular}
\end{table}

\begin{table}
\caption{Values for $\kappa_n$ and $\kappa_s$ determined for three
different quantities $Q$ to set the lattice scale. All estimates were
obtained by setting $\bm$ equal to its tadpole improved perturbative
value. 
\label{tab_kappa_phys}}
\begin{tabular}{lcc cc c cc c cc}
 &         &                & \multicolumn{2}{c}{$Q=r_0^{-1}$} &
                            & \multicolumn{2}{c}{$Q=m_{K^*}$}  &
                            & \multicolumn{2}{c}{$Q=m_{N}$}   \\
 & $\beta$ & $L^3\cdot T$   & $\kappa_n$ & $\kappa_s$ &&
   $\kappa_n$ & $\kappa_s$ && $\kappa_n$ & $\kappa_s$ \\
\tableline
NP  & 6.0  & $16^3\cdot48$ 
           & 0.13520$^{+1}_{-2}$ & 0.13401$^{+2}_{-2}$ &&
             0.13520$^{+2}_{-1}$ & 0.13383$^{+4}_{-6}$ &&
             0.13517$^{+2}_{-2}$ & 0.13318$^{+19}_{-18}$ \\
    & 6.0  & $32^3\cdot64$
           & 0.13519$^{+1}_{-1}$ & 0.13398$^{+2}_{-2}$ &&
             0.13518$^{+1}_{-1}$ & 0.13376$^{+5}_{-7}$ &&
             0.13516$^{+1}_{-1}$ & 0.13327$^{+19}_{-14}$ \\
    & 6.2  & $24^3\cdot48$
           & 0.13578$^{+1}_{-2}$ & 0.13495$^{+2}_{-2}$ &&
             0.13577$^{+2}_{-1}$ & 0.13476$^{+3}_{-5}$ &&
             0.13576$^{+2}_{-1}$ & 0.13438$^{+13}_{-~4}$ \\
\tableline
TAD & 5.7  & $16^3\cdot48$ 
           & 0.14306$^{+4}_{-3}$ & 0.14012$^{+5}_{-3}$ &&
             0.14307$^{+3}_{-4}$ & 0.14029$^{+11}_{-10}$ &&
             0.14302$^{+3}_{-4}$ & 0.13904$^{+22}_{-21}$ \\
    & 6.0  & $16^3\cdot48$ 
           & 0.13915$^{+1}_{-2}$ & 0.13780$^{+1}_{-2}$ &&
             0.13915$^{+2}_{-1}$ & 0.13769$^{+5}_{-5}$ &&
             0.13913$^{+2}_{-1}$ & 0.13712$^{+13}_{-20}$ \\
    & 6.2  & $24^3\cdot48$
           & 0.13786$^{+1}_{-2}$ & 0.13699$^{+2}_{-2}$ &&
             0.13785$^{+2}_{-1}$ & 0.13680$^{+3}_{-5}$ &&
             0.13784$^{+2}_{-1}$ & 0.13655$^{+12}_{-~6}$ \\
\end{tabular}
\end{table}

\begin{table}
\caption{Results for the physical meson and baryon masses with the
lattice scale set by $r_0$.
\label{tab_res_r0}}
\begin{tabular}{c cc c cc c cc c cc}
%\begin{tabular}{c cc c cc c cc}
  & \multicolumn{2}{c}{$m_\rho\,r_0$}  &
  & \multicolumn{2}{c}{$m_{K^*}\,r_0$} &
  & \multicolumn{2}{c}{$m_\phi\,r_0$}  &
  & \multicolumn{2}{c}{$\mbox{}$}  \\
$\beta$ & NP & TAD && NP & TAD && NP & TAD && & \\
\tableline
5.7  &                     & 1.982$^{+38}_{-54}$ && 
                           & 2.224$^{+28}_{-39}$ &&
                           & 2.466$^{+19}_{-25}$ &&
                           &                     \\
6.0  & 2.189$^{+56}_{-56}$ & 2.121$^{+48}_{-49}$ &&
       2.422$^{+41}_{-42}$ & 2.356$^{+36}_{-36}$ &&
       2.655$^{+29}_{-28}$ & 2.592$^{+25}_{-23}$ &&
                           &                     \\
6.2  & 2.254$^{+69}_{-81}$ & 2.211$^{+66}_{-67}$ &&
       2.468$^{+51}_{-55}$ & 2.433$^{+48}_{-50}$ &&
       2.682$^{+34}_{-35}$ & 2.654$^{+33}_{-34}$ &&
                           &                     \\
Cont.& \multicolumn{2}{c}{2.352(163)}        &&
       \multicolumn{2}{c}{2.540(117)}        &&
       \multicolumn{2}{c}{2.729(77)}         &&
       \multicolumn{2}{c}{$\mbox{}$}         \\
\tableline
\tableline
  & \multicolumn{2}{c}{$m_N\,r_0$}  &
  & \multicolumn{2}{c}{$m_{\Sigma}\,r_0$} &
  & \multicolumn{2}{c}{$m_{\Lambda}\,r_0$} &
  & \multicolumn{2}{c}{$m_{\Xi}\,r_0$}  \\
$\beta$ & NP & TAD && NP & TAD && NP & TAD && NP & TAD \\
\tableline
5.7 &                    & 2.74$^{+~6}_{-~7}$ &&
                         & 3.05$^{+~5}_{-~5}$ &&
                         & 3.05$^{+~5}_{-~5}$ &&
                         & 3.35$^{+~5}_{-~4}$ \\
6.0 & 3.08$^{+13}_{-13}$ & 2.92$^{+14}_{-~9}$ &&
      3.36$^{+10}_{-11}$ & 3.22$^{+11}_{-~7}$ &&
      3.34$^{+11}_{-11}$ & 3.21$^{+11}_{-~7}$ &&
      3.63$^{+~9}_{-~8}$ & 3.52$^{+~9}_{-~5}$ \\
6.2 & 3.02$^{+~4}_{-14}$ & 2.87$^{+~6}_{-12}$ &&
      3.31$^{+~4}_{-11}$ & 3.18$^{+~5}_{-~9}$ &&
      3.30$^{+~3}_{-12}$ & 3.18$^{+~4}_{-10}$ &&
      3.59$^{+~3}_{-~8}$ & 3.50$^{+~4}_{-~7}$ \\
Cont.& \multicolumn{2}{c}{2.92(24)}        &&
       \multicolumn{2}{c}{3.23(19)}        &&
       \multicolumn{2}{c}{3.22(20)}        &&
       \multicolumn{2}{c}{3.54(15)}        \\
\tableline
\tableline
  & \multicolumn{2}{c}{$m_{\Delta}\,r_0$}  &
  & \multicolumn{2}{c}{$m_{\Sigma^*}\,r_0$} &
  & \multicolumn{2}{c}{$m_{\Xi^*}\,r_0$}  &
  & \multicolumn{2}{c}{$m_{\Omega}\,r_0$}  \\
$\beta$ & NP & TAD && NP & TAD && NP & TAD && NP & TAD \\
\tableline
5.7 &                    & 3.29$^{+12}_{-~9}$ &&
                         & 3.55$^{+10}_{-~7}$ &&
                         & 3.81$^{+~8}_{-~6}$ &&
                         & 4.08$^{+~7}_{-~4}$ \\
6.0 & 4.12$^{+26}_{-19}$ & 3.93$^{+11}_{-16}$ &&
      4.29$^{+20}_{-16}$ & 4.13$^{+~9}_{-13}$ &&
      4.47$^{+14}_{-12}$ & 4.32$^{+~6}_{-10}$ &&
      4.64$^{+~9}_{-~9}$ & 4.52$^{+~5}_{-~8}$ \\
6.2 & 4.02$^{+16}_{-12}$ & 3.96$^{+11}_{-~9}$ &&
      4.24$^{+13}_{-10}$ & 4.18$^{+~9}_{-~7}$ &&
      4.46$^{+10}_{-~8}$ & 4.40$^{+~7}_{-~6}$ &&
      4.68$^{+~7}_{-~6}$ & 4.62$^{+~6}_{-~5}$ \\
Cont.& \multicolumn{2}{c}{3.86(37)}        &&
       \multicolumn{2}{c}{4.15(29)}        &&
       \multicolumn{2}{c}{4.44(22)}        &&
       \multicolumn{2}{c}{4.72(17)}        \\
\end{tabular}
\end{table}

\begin{table}
\caption{Same as Table~\protect\ref{tab_res_r0} but with the scale set
by $m_{K^*}$.
\label{tab_res_mkstar}}
\begin{tabular}{c cc c cc c cc c cc}
%\begin{tabular}{c cc c cc c cc}
  & \multicolumn{2}{c}{$m_\rho/m_{K^*}$}  &
  & \multicolumn{2}{c}{$m_\phi/m_{K^*}$}  &
  & \multicolumn{2}{c}{$\mbox{}$}  &
  & \multicolumn{2}{c}{$\mbox{}$}  \\
$\beta$ & NP & TAD && NP & TAD && & && & \\
\tableline
5.7  &                     & 0.896$^{+~3}_{-~9}$ &&
                           & 1.105$^{+~2}_{-~4}$ &&
                           &                     &&
                           &                     \\
6.0  & 0.890$^{+10}_{-17}$ & 0.887$^{+10}_{-13}$ &&
       1.113$^{+~1}_{-~7}$ & 1.107$^{+~4}_{-~6}$ &&
                           &                     &&
                           &                     \\
6.2  & 0.903$^{+12}_{-25}$ & 0.893$^{+14}_{-23}$ &&
       1.112$^{+~2}_{-10}$ & 1.107$^{+~2}_{-10}$ &&
                           &                     &&
                           &                     \\
Cont.& \multicolumn{2}{c}{0.921$^{+32}_{-56}$}   &&
       \multicolumn{2}{c}{1.110$^{+~8}_{-21}$}   &&
       \multicolumn{2}{c}{$\mbox{}$}         &&
       \multicolumn{2}{c}{$\mbox{}$}         \\
\tableline
\tableline
  & \multicolumn{2}{c}{$m_N/m_{K^*}$}  &
  & \multicolumn{2}{c}{$m_{\Sigma}/m_{K^*}$} &
  & \multicolumn{2}{c}{$m_{\Lambda}/m_{K^*}$} &
  & \multicolumn{2}{c}{$m_{\Xi}/m_{K^*}$}  \\
$\beta$ & NP & TAD && NP & TAD && NP & TAD && NP & TAD \\
\tableline
5.7 &                     & 1.239$^{+37}_{-37}$ &&
                          & 1.372$^{+32}_{-29}$ &&
                          & 1.372$^{+32}_{-29}$ &&
                          & 1.504$^{+28}_{-22}$ \\
6.0 & 1.253$^{+51}_{-64}$ & 1.220$^{+57}_{-41}$ &&
      1.385$^{+41}_{-50}$ & 1.360$^{+46}_{-32}$ &&
      1.379$^{+40}_{-50}$ & 1.356$^{+45}_{-31}$ &&
      1.517$^{+31}_{-35}$ & 1.500$^{+36}_{-24}$ \\
6.2 & 1.212$^{+18}_{-68}$ & 1.160$^{+22}_{-56}$ &&
      1.352$^{+15}_{-52}$ & 1.312$^{+18}_{-43}$ &&
      1.347$^{+13}_{-55}$ & 1.313$^{+13}_{-44}$ &&
      1.491$^{+14}_{-37}$ & 1.465$^{+13}_{-33}$ \\
Cont.& \multicolumn{2}{c}{1.14$^{+~6}_{-18}$}        &&
       \multicolumn{2}{c}{1.29$^{+~5}_{-14}$}        &&
       \multicolumn{2}{c}{1.29$^{+~5}_{-15}$}        &&
       \multicolumn{2}{c}{1.45$^{+~4}_{-10}$}        \\
\tableline
\tableline
  & \multicolumn{2}{c}{$m_{\Delta}/m_{K^*}$}  &
  & \multicolumn{2}{c}{$m_{\Sigma^*}/m_{K^*}$} &
  & \multicolumn{2}{c}{$m_{\Xi^*}/m_{K^*}$}  &
  & \multicolumn{2}{c}{$m_{\Omega}/m_{K^*}$}  \\
$\beta$ & NP & TAD && NP & TAD && NP & TAD && NP & TAD \\
\tableline
5.7 &                     & 1.487$^{+55}_{-46}$ &&
                          & 1.600$^{+49}_{-36}$ &&
                          & 1.713$^{+43}_{-28}$ &&
                          & 1.826$^{+36}_{-21}$ \\
6.0 & 1.672$^{+99}_{-88}$ & 1.645$^{+44}_{-71}$ &&
      1.756$^{+70}_{-67}$ & 1.735$^{+34}_{-60}$ &&
      1.841$^{+46}_{-52}$ & 1.825$^{+26}_{-47}$ &&
      1.926$^{+26}_{-40}$ & 1.916$^{+20}_{-37}$ \\
6.2 & 1.609$^{+59}_{-58}$ & 1.598$^{+42}_{-47}$ &&
      1.716$^{+43}_{-50}$ & 1.705$^{+32}_{-38}$ &&
      1.823$^{+31}_{-41}$ & 1.811$^{+24}_{-32}$ &&
      1.930$^{+22}_{-36}$ & 1.918$^{+20}_{-30}$ \\
Cont.& \multicolumn{2}{c}{1.50$^{+17}_{-17}$}        &&
       \multicolumn{2}{c}{1.64$^{+13}_{-13}$}        &&
       \multicolumn{2}{c}{1.79$^{+~9}_{-10}$}        &&
       \multicolumn{2}{c}{1.93$^{+~7}_{-~8}$}        \\
\end{tabular}
\end{table}

\begin{table}
\caption{Same as Table~\protect\ref{tab_res_r0} but with the scale set
by the nucleon mass $m_N$.
\label{tab_res_mnuc}}
\begin{tabular}{c cc c cc c cc c cc}
%\begin{tabular}{c cc c cc c cc}
  & \multicolumn{2}{c}{$m_\rho/m_N$}  &
  & \multicolumn{2}{c}{$m_{K^*}/m_N$}  &
  & \multicolumn{2}{c}{$m_\phi/m_N$}  &
  & \multicolumn{2}{c}{$\mbox{}$}  \\
$\beta$ & NP & TAD && NP & TAD && NP & TAD && & \\
\tableline
5.7  &                     & 0.727$^{+25}_{-30}$ &&
                           & 0.843$^{+18}_{-20}$ &&
                           & 0.959$^{+13}_{-12}$ &&
                           &                     \\
6.0  & 0.709$^{+37}_{-34}$ & 0.726$^{+27}_{-36}$ &&
       0.837$^{+28}_{-24}$ & 0.848$^{+20}_{-27}$ &&
       0.965$^{+21}_{-17}$ & 0.970$^{+15}_{-20}$ &&
                           &                     \\
6.2  & 0.744$^{+44}_{-23}$ & 0.769$^{+39}_{-21}$ &&
       0.859$^{+34}_{-14}$ & 0.882$^{+31}_{-15}$ &&
       0.975$^{+26}_{-~7}$ & 0.995$^{+24}_{-12}$ &&
                           &                     \\
Cont.& \multicolumn{2}{c}{0.796$^{+86}_{-48}$}   &&
       \multicolumn{2}{c}{0.894$^{+65}_{-29}$}   &&
       \multicolumn{2}{c}{0.992$^{+51}_{-22}$}   &&
       \multicolumn{2}{c}{$\mbox{}$}         \\
\tableline
\tableline
  & \multicolumn{2}{c}{$m_{\Sigma}/m_N$} &
  & \multicolumn{2}{c}{$m_{\Lambda}/m_N$} &
  & \multicolumn{2}{c}{$m_{\Xi}/m_N$}  &
  & \multicolumn{2}{c}{$\mbox{}$}  \\
$\beta$ & NP & TAD && NP & TAD && NP & TAD && &  \\
\tableline
5.7 &                     & 1.150$^{+~5}_{-~3}$ &&
                          & 1.150$^{+~5}_{-~3}$ &&
                          & 1.299$^{+~9}_{-~5}$ &&
                          &                     \\
6.0 & 1.151$^{+10}_{-~8}$ & 1.156$^{+~6}_{-~7}$ &&
      1.148$^{+10}_{-~9}$ & 1.153$^{+~7}_{-~9}$ &&
      1.303$^{+20}_{-16}$ & 1.311$^{+11}_{-13}$ &&
                          &                     \\
6.2 & 1.154$^{+12}_{-~6}$ & 1.162$^{+~8}_{-~5}$ &&
      1.151$^{+11}_{-~9}$ & 1.162$^{+10}_{-10}$ &&
      1.308$^{+23}_{-12}$ & 1.323$^{+16}_{-~9}$ &&
                          &                     \\
Cont.& \multicolumn{2}{c}{1.161$^{+23}_{-12}$}  &&
       \multicolumn{2}{c}{1.157$^{+21}_{-20}$}  &&
       \multicolumn{2}{c}{1.321$^{+46}_{-25}$}  &&
       \multicolumn{2}{c}{$\mbox{}$}        \\
\tableline
\tableline
  & \multicolumn{2}{c}{$m_{\Delta}/m_N$}  &
  & \multicolumn{2}{c}{$m_{\Sigma^*}/m_N$} &
  & \multicolumn{2}{c}{$m_{\Xi^*}/m_N$}  &
  & \multicolumn{2}{c}{$m_{\Omega}/m_N$}  \\
$\beta$ & NP & TAD && NP & TAD && NP & TAD && NP & TAD \\
\tableline
5.7 &                     & 1.198$^{+49}_{-40}$ &&
                          & 1.326$^{+42}_{-30}$ &&
                          & 1.454$^{+35}_{-23}$ &&
                          & 1.582$^{+31}_{-16}$ \\
6.0 & 1.330$^{+99}_{-75}$ & 1.344$^{+47}_{-83}$ &&
      1.427$^{+76}_{-60}$ & 1.445$^{+37}_{-70}$ &&
      1.524$^{+53}_{-51}$ & 1.545$^{+28}_{-57}$ &&
      1.621$^{+49}_{-57}$ & 1.646$^{+28}_{-49}$ \\
6.2 & 1.325$^{+94}_{-36}$ & 1.376$^{+78}_{-35}$ &&
      1.443$^{+76}_{-30}$ & 1.488$^{+65}_{-28}$ &&
      1.561$^{+60}_{-23}$ & 1.601$^{+57}_{-25}$ &&
      1.679$^{+49}_{-24}$ & 1.714$^{+50}_{-23}$ \\
Cont.& \multicolumn{2}{c}{1.33$^{+20}_{-10}$}        &&
       \multicolumn{2}{c}{1.47$^{+16}_{-~7}$}        &&
       \multicolumn{2}{c}{1.61$^{+12}_{-~5}$}        &&
       \multicolumn{2}{c}{1.76$^{+10}_{-~4}$}        \\
\end{tabular}
\end{table}

\begin{table}
\caption{Results for the physical meson and baryons masses at
$\beta=6.0$ on $32^3\cdot64$.
\label{tab_all_res_B60LARGE}}
\begin{tabular}{l r@{.}l r@{.}l r@{.}l r@{.}l}
  & \multicolumn{2}{c}{$m_\rho/Q$}  
  & \multicolumn{2}{c}{$m_{K^*}/Q$} 
  & \multicolumn{2}{c}{$m_\phi/Q$}  
  & \multicolumn{2}{c}{$\mbox{}$}  \\
\tableline
$Q=r_0^{-1}$ & 
   2&243$^{+59}_{-51}$ & 2&461$^{+45}_{-38}$ &
   2&680$^{+32}_{-27}$ & \multicolumn{2}{c}{$\mbox{}$}  \\
$Q=m_{K^*}$  & 
   0&898$^{+13}_{-12}$ & \multicolumn{2}{c}{$\mbox{}$} &
   1&111$^{+~5}_{-~7}$ & \multicolumn{2}{c}{$\mbox{}$}  \\
$Q=m_{N}$    & 
   0&746$^{+42}_{-27}$ & 0&863$^{+34}_{-21}$ &
   0&980$^{+28}_{-17}$ & \multicolumn{2}{c}{$\mbox{}$}  \\
\tableline
\tableline
  & \multicolumn{2}{c}{$m_N/Q$}  
  & \multicolumn{2}{c}{$m_{\Sigma}/Q$} 
  & \multicolumn{2}{c}{$m_{\Lambda}/Q$}
  & \multicolumn{2}{c}{$m_{\Xi}/Q$}  \\
\tableline
$Q=r_0^{-1}$ & 
   3&00$^{+10}_{-15}$ & 3&29$^{+~8}_{-11}$ &
   3&27$^{+~8}_{-12}$ & 3&58$^{+~7}_{-~9}$ \\
$Q=m_{K^*}$  & 
   1&201$^{+43}_{-62}$ & 1&343$^{+32}_{-51}$ &
   1&337$^{+32}_{-52}$ & 1&484$^{+25}_{-38}$ \\
$Q=m_{N}$    & 
   \multicolumn{2}{c}{$\mbox{}$} & 1&155$^{+~9}_{-~9}$ &
   1&151$^{+~8}_{-~9}$ & 1&310$^{+18}_{-18}$ \\
\tableline
\tableline
  & \multicolumn{2}{c}{$m_{\Delta}/Q$} 
  & \multicolumn{2}{c}{$m_{\Sigma^*}/Q$}
  & \multicolumn{2}{c}{$m_{\Xi^*}/Q$}  
  & \multicolumn{2}{c}{$m_{\Omega}/Q$}  \\
\tableline
$Q=r_0^{-1}$ & 
   3&81$^{+14}_{-10}$ & 4&03$^{+12}_{-~8}$ &
   4&26$^{+10}_{-~8}$ & 4&48$^{+~8}_{-~7}$ \\
$Q=m_{K^*}$  & 
   1&525$^{+58}_{-46}$ & 1&635$^{+46}_{-41}$ &
   1&745$^{+38}_{-37}$ & 1&855$^{+32}_{-37}$ \\
$Q=m_{N}$    & 
   1&267$^{+73}_{-43}$ & 1&387$^{+63}_{-37}$ & 
   1&508$^{+53}_{-33}$ & 1&628$^{+46}_{-34}$ \\
\end{tabular}
\end{table}

\begin{table}
\caption{Results for vector mesons, octet ($\Sigma$-like) and decuplet
($\Delta$-like) baryons, in units of $r_0$, interpolated to the
reference points defined by $(\mps{r_0})^2=3.0$ and
$m_{\rm{PS}}/m_{\rm{V}}=0.7$. 
\label{tab_res_mref}}
\begin{tabular}{c cc c cc c cc}
  & \multicolumn{8}{c}{$(\mps{r_0})^2=3.0$} \\
  & \multicolumn{2}{c}{$\mv\,r_0$}  &
  & \multicolumn{2}{c}{$m_{\Sigma}\,r_0$}  &
  & \multicolumn{2}{c}{$m_{\Delta}\,r_0$}  \\
$\beta$ & NP & TAD && NP & TAD && NP & TAD \\
\tableline
5.7  &                     & 2.466$^{+23}_{-31}$ &&
                           & 3.651$^{+39}_{-28}$ &&
                           & 4.067$^{+69}_{-40}$ \\
6.0  & 2.649$^{+29}_{-29}$ & 2.586$^{+25}_{-23}$ &&
       3.897$^{+67}_{-54}$ & 3.805$^{+70}_{-41}$ &&
       4.638$^{+89}_{-93}$ & 4.508$^{+44}_{-79}$ \\
6.2  & 2.677$^{+34}_{-35}$ & 2.649$^{+33}_{-34}$ &&
       3.871$^{+36}_{-63}$ & 3.806$^{+36}_{-56}$ &&
       4.668$^{+70}_{-65}$ & 4.612$^{+57}_{-49}$ \\
Cont.& \multicolumn{2}{c}{2.725(78)}   &&
       \multicolumn{2}{c}{3.83(12)}    &&
       \multicolumn{2}{c}{4.71(17)}    \\
\tableline
\tableline
  & \multicolumn{8}{c}{$\mps/\mv=0.7$} \\
  & \multicolumn{2}{c}{$\mv\,r_0$}  &
  & \multicolumn{2}{c}{$m_{\Sigma}\,r_0$}  &
  & \multicolumn{2}{c}{$m_{\Delta}\,r_0$}  \\
$\beta$ & NP & TAD && NP & TAD && NP & TAD \\
\tableline
5.7  &                     & 2.451$^{+32}_{-35}$ &&
                           & 3.633$^{+49}_{-41}$ &&
                           & 4.051$^{+79}_{-55}$ \\
6.0  & 2.760$^{+37}_{-35}$ & 2.669$^{+37}_{-25}$ &&
       4.112$^{+68}_{-48}$ & 3.957$^{+75}_{-37}$ &&
       4.776$^{+64}_{-80}$ & 4.607$^{+46}_{-69}$ \\
6.2  & 2.796$^{+45}_{-39}$ & 2.758$^{+43}_{-36}$ &&
       4.115$^{+63}_{-66}$ & 4.043$^{+53}_{-56}$ &&
       4.855$^{+63}_{-68}$ & 4.777$^{+62}_{-52}$ \\
Cont.& \multicolumn{2}{c}{2.845(94)}   &&
       \multicolumn{2}{c}{4.13(15)}    &&
       \multicolumn{2}{c}{4.97(15)}    \\
\end{tabular}
\end{table}

%%%%%%%%%%%%%%%%%%%%%%%%%%%%%%%%%%%%%%%%%%%%%%%%%%%%%%%%%%%%%%%%%%%%%%%

\end{document}